\documentclass[prb,twocolumn,superscriptaddress,floatfix,graphicx,showpacs]{revtex4-1}
\usepackage{ifpdf}
 \newif\ifpdf
\ifx\pdfoutput\undefined
   \pdffalse
\else
   \pdfoutput=1
   \pdftrue
\fi
\ifpdf
   \usepackage{graphicx}
   \usepackage{epstopdf}
   \usepackage{color}
\else
   \usepackage{graphicx}
\fi
\usepackage{cancel}
\usepackage{srctex}
\usepackage{hyperref}

\usepackage{amsmath,amssymb}

\usepackage{epsfig}

\usepackage{epsfig}

\begin{document}

\title{Structure and glass-forming ability of simulated Ni-Zr alloys}

\author{B.A. Klumov}
\affiliation{High Temperature Institute, Russian Academy of Sciences, 125412, Moscow, Russia}
\affiliation{L.D. Landau Institute for Theoretical Physics, Russian Academy of Sciences, 117940 Moscow, Russia}
\affiliation{Ural Federal University, 620002 Ekaterinburg, Russia}

\author{R.E. Ryltsev}
\affiliation{Institute of Metallurgy, Ural Branch of Russian Academy of Sciences, 620016 Ekaterinburg, Russia}
\affiliation{Ural Federal University,620002 Ekaterinburg, Russia}
\affiliation{L.D. Landau Institute for Theoretical Physics, Russian Academy of Sciences, 117940 Moscow, Russia}

\author{N.M. Chtchelkatchev}
\affiliation{L.D. Landau Institute for Theoretical Physics, Russian Academy of Sciences, 117940 Moscow, Russia}
\affiliation{Institute of Metallurgy, Ural Branch of Russian Academy of Sciences, 620016 Ekaterinburg, Russia}
\affiliation{Moscow Institute of Physics and Technology, 141700 Moscow, Russia}
\affiliation{Institute for High Pressure Physics, Russian Academy of Sciences, 108840 Moscow (Troitsk), Russia}


\begin{abstract}
Binary Cu-Zr system is a representative bulk glassformer demonstrating high glass forming ability due to pronounced icosahedral local ordering. From the first glance, Ni-Zr system is the most natural object to expect the same behavior because nickel and copper are neighbours in the periodic table and have similar physicochemical properties. However, doing molecular dynamics simulations of $\rm Ni_{\alpha}Zr_{1-\alpha}$ alloys described by embedded atom model potential, we observe different behaviour. We conclude that the Ni-Zr system  has the same glass-forming ability as an additive binary Lennard-Jones mixture without any chemical interaction. The structural analysis reveals that icosahedral ordering in Ni-Zr alloys is much less pronounced than that in the Cu-Zr ones. We suggest that lack of icosahedral ordering due to peculiarities of interatomic interactions is the reason of relatively poor glass-forming ability of Ni-Zr system.
\end{abstract}

\maketitle

\section{Introduction}
Metallic glasses, solid metallic materials disordered at atomic scale, are of special interest due to their extraordinary physical properties and high potential for applications \cite{Suryanarayana2017BMG_Book,Smith2017NaturePhys,Hufnagel2016ActaMater,Laws2015NatureComm}. Important problem is identifying alloys with high glass-forming ability (GFA), which are capable to form metallic glasses under cooling of melts at relatively low rates \cite{Wu2015NatureComm,Yu2015JAlloysComp}. Empirical knowledge of GFA is mostly based on statistical guides and there is a lack of microscopic rationale~\cite{Tang2013Nature}. A number of good metallic glass-forming alloys is known, with Cu-Zr ones as an exemplary glassformers capable to form bulk metallic glasses ~\cite{Xu2004ActMat,Wang2004AppPhysLett,Wang2005JMatRes}.  Cu and Zr are both d-elements in the periodic table with pronounced metallic properties. What happens if we replace one of the components on a ``relative'' metallic d-element? This is an example of one of the most important issues, the solution of which is necessary to understand the nature of metallic glasses.  We address this question moving from Cu-Zr to Ni-Zr alloys.

Cu and Ni are neighbors in the periodic table, with atomic numbers 29 and 28 and atomic weights 63.54 and 58.69, respectively. These elements are completely miscible in both liquid and solid states forming fcc-based Cu-Ni alloys which are widely used (e.g., in coinage). So the comparison of Cu-Zr and Ni-Zr alloys seems the most natural problem.

The relation between local structure and GFA of metallic alloys is an important issue from both fundamental and applied points of view~\cite{Royall2015PhysRep}. One of the widely accepted paradigm is that local icosahedral ordering is the cause of geometrical frustration favoring GFA~\cite{Tarjus2005JPCM,Berthier2011RevModPhys,Shintani2006NaturePhys}. It is widely accepted by both experiments and simulations that GFA of Cu-Zr-based alloys is caused by pronounced icosahedral local ordering~\cite{Li2009PRB,Cheng2009PRL,Soklaski2013PRB,Wu2013PRB,Wen2013JNonCrystSol,Wang2015JPhysChemA,Wu2013PRB,Saksl2003ApplPhysLett}. Note that importance of other types of local clusters has been also pointed out~\cite{HM2012, Peng2010ApplPhysLett}. There is a natural question if the glass-forming properties of Cu-Zr system are universal for all similar alloys. Experimentally, the GFA of Ni-Zr system is essentially lower than that for Cu-Zr one and so bulk metallic glasses can not be produced for Ni-Zr alloys \cite{Dong1981JNCS,Huang2013JM}. Thus, the comparison of structural characteristics of Cu-Zr and Ni-Zr alloys is an important fundamental task to understand the striking difference between the GFA. Experimental methods of structural analysis, even together with  reverse Monte-Carlo technique, do not give complete picture and reveal qualitatively different results~\cite{Holland-Moritz2009PRB,Johnson2014JNCS,Fukunaga2006Intermet,Kaban2013, Liu2009PhysLettA,Liu2008ApplPhysLett}.  {\it Ab-initio} molecular dynamic simulations can not completely resolve these controversies \cite{Hao2009PRB} possibly because of the limited amount of atoms in the simulated systems and quite small (picosecond) simulation times, not enough for reliable simulation of supercooled liquids and glasses. Here we address this issue doing classical molecular dynamic simulations, free from most of these limitations. We use Ni-Zr embedded atom model (EAM) potential proposed in \cite{Wilson2015PhilosMag}. Its reliability is ensured by good agreement with experimental data and quantum simulations of liquid Ni-Zr alloys.

We have done the structural analysis of Ni-Zr. It reveals that icosahedral ordering in Ni-Zr alloys is much less pronounced than that for the Cu-Zr ones. We show that simulated Ni-Zr system has rather poor GFA comparable to that for additive binary Lennard-Jones mixture without any chemical interaction (that means  potential with
$\sigma_{\scriptscriptstyle{\rm AB}}=(\sigma_{\scriptscriptstyle{\rm AA}} +\sigma_{\scriptscriptstyle{\rm BB}})/2$ and $\epsilon_{\scriptscriptstyle{\rm AA}}=\epsilon_{\scriptscriptstyle{\rm BB}}=\epsilon_{\scriptscriptstyle{\rm BB}}$)\cite{Shimono2001ScriptaMater,Shimono2012RevueMet}.

The Ni-Zr system demonstrates different solid phases, both ordered and disordered, being quenched at cooling rates of the order of $10^{10}-10^{12}$ K/s usually applied in molecular dynamic simulations. So it is a good model for studying the interplay between crystallization and glass formation.

Doing such studies one should have reliable methods for detecting both crystalline and glassy states as well as recognizing short-ranged structures with different symmetries. Thus, the purpose of this paper is twofold: to study the GFA of simulated Ni-Zr alloys in the whole range of the compositions, and to develop new instruments for analysing the structure of quenched solid phases complementing the known methods.

\section{Methods\label{sec_methods}}
Classical molecular dynamics is the main theoretical tool to study properties of glasses because it makes it possible to overcome the problems of analytical description of non-ordered condensed matter systems \cite{Liu1987PRB, Dubinin2014RussianChemRev,Dubinin2011ThermActa} and allows studying microscopic structure and dynamics covering sufficiently large time and spatial scales \cite{Frenkel2001Textbook}.

For the molecular dynamics (MD) simulations, we use LAMMPS package \cite{Plimpton1995JCompPhys}. The system of $N \approx 5000$ particles was simulated under periodic boundary conditions in Nose-Hoover NPT ensemble at pressure $P=1$~atm. The MD time step was 1-3~fs depending on the system temperature. 
Initial configurations were equilibrated melts at $T = 1800$~K. Then, the system was cooled down to $T=300$~K with different cooling rates: $\gamma \in (10^{10}, 10^{12})$~K/s.

As the model of interaction between alloy components, we use EAM potential of Finnis-Sinclair type developed in \cite{Wilson2015PhilosMag}.

To identify the crystalline and liquid-like particles we use the bond orientational order parameters method (BOOP)~\cite{Steinhardt1983}; the method has been widely used in condensed matter physics to quantify the local orientational order~\cite{Steinhardt1981,Steinhardt1983,Mitus1982} in Lennard-Jones and hard sphere systems~\cite{RT96,Luchnikov2002,Jin2010,Ryltsev2013PRE,KLumovJPC14,Bar14}, bulky and confined complex plasmas~\cite{Klumov2007JETPLett,Klumov2010EPL,Klumov2010UFN,Khrapak2012}, colloidal~\cite{Gasser01,Kawasaki10,Ryltsev2013PRL,Ryltsev2017SoftMatt} and patchy systems~\cite{Vasilyev2013,Vasilyev2015}, etc. The method allows us to explicitly recognize symmetry of the local atomic clusters~\cite{KlumovPU10,KlumovPRB11,Hirata2013Science,KLumovJPC14} and study their spatial distribution~\cite{Ryltsev2013PRE,Ryltsev2015SoftMatt,Ryltsev2016JCP}.

Within the framework of BOOP method, we define the rotational invariants (RI) of rank $l$ of both the second $q_l$ and third order $w_l$~\cite{Steinhardt1981,Steinhardt1983}. The advantage of $q_l$ and $w_l$ is that they are uniquely determined for any polyhedron including the elements of any crystalline structure. Among the RI, $q_4$, $q_6$, $w_4$, $w_6$ are typically the most informative ones so we use them in this study.
To identify close packed and icosahedral-like clusters we calculate the rotational invariants $q_l$, $w_l$ for each atom using the fixed number of nearest neighbors ($N_{\rm nn}=12$). Atom whose coordinates in $(q_4, q_6, w_4, w_6)$ space are sufficiently close to those for the perfect structures are counted as icosahedral-like (fcc-like, hcp-like) etc. The RI for a number of close-packed structures are shown in Table~\ref{table1}.

\begin{table}
\centering
\caption{Rotational invariants (RI) $q_l$ and $w_l$ ($l=4,~6$) of a few
perfect clusters calculated via fixed number of nearest neighbors (NN):
hexagonal close-packed (hcp), face centered cubic (fcc), icosahedron (ico), body-centered cubic (bcc) (calculated for both first (NN = 8) and second (NN=14) shells). Additionally, mean RI for the Lennard-Jones melt are shown for the comparison.}
\begin{tabular}{|c|c|c|c|c|}
\hline 
cluster  & \quad $q_{4}$ & \quad $q_{6}$ & \quad $w_{4}$ & \quad $ w_{6}$ \\ \hline
hcp (12 NN) & 0.097 & 0.485 & 0.134  & -0.012 \\ \hline
fcc (12 NN) & 0.19  & 0.575  & -0.159  &  -0.013 \\ \hline
ico (12 NN) & $1.4 \times 10^{-4}$ & 0.663 & -0.159  & -0.169 \\ \hline
bcc ( 8 NN) & 0.5 & 0.628 & -0.159   & 0.013 \\ \hline
bcc (14 NN) & 0.036 & 0.51 & 0.159   & 0.013 \\ \hline
LJ melt (12 NN) & $\approx $0.155  & $\approx $0.37  & $\approx $-0.023  &$\approx $-0.04 \\ \hline
\end{tabular}
\label{table1}
\end{table}

\begin{figure}
\centering
\includegraphics[width=0.7\columnwidth]{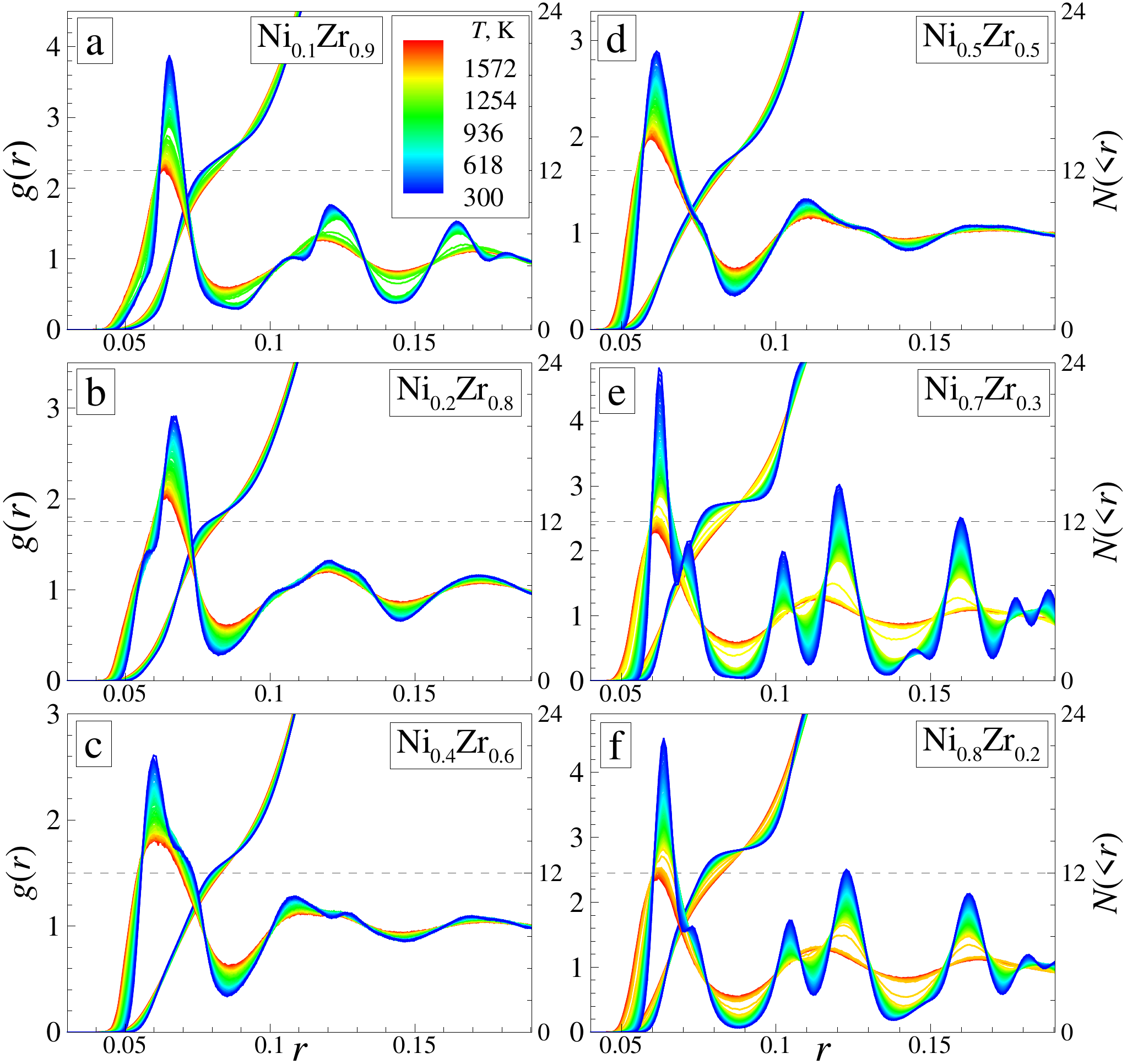}\\
\caption{(Color online) Total radial distribution functions $g(r)$ of ${\rm Ni_{\alpha}Zr_{1-\alpha}}$ alloys quenched from the liquid state (equilibrated at temperature $T$ = 1800 K) down to 300 K. Temperature evolution of $g(r)$ is shown at different Ni abundance $\alpha$ (indicated on each panel). The curves are color-coded by temperature $T$. The cumulative measure $N(<r)=4\pi\rho\int_0^r{\xi^2g(\xi)d\xi}$ is also plotted to evaluate the nearest neighbours number $N_{\rm n}$ in the first coordination shell (grey dashed line corresponds to the closely packed structures ($N_{\rm n}$=12)). The cooling rate $\gamma$  is $10^{11}$ K/s.}
\label{fig:RDF_tot_vs_T}
\end{figure}

\begin{figure}
\centering
\includegraphics[width=0.7\columnwidth]{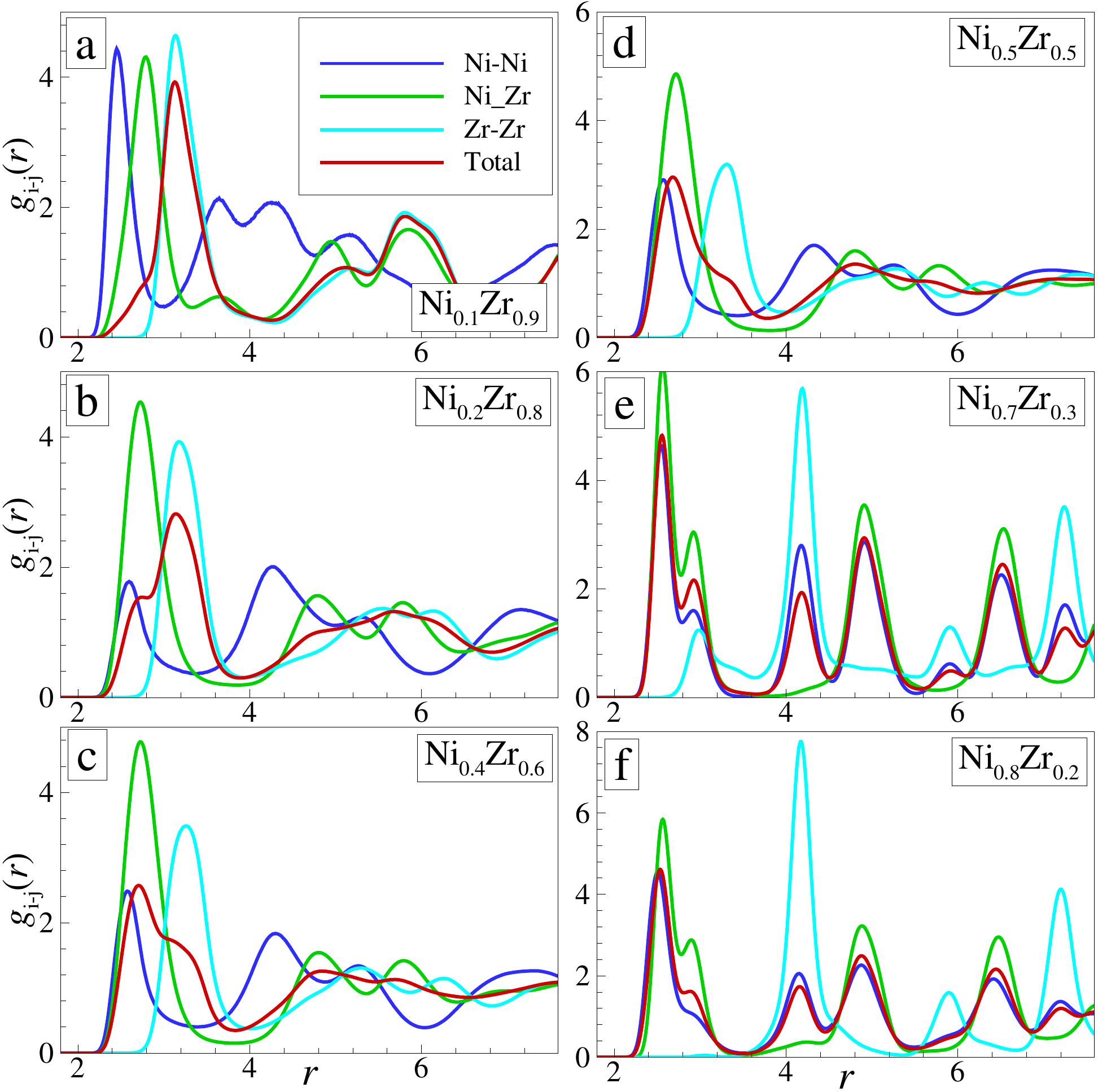}\\
\caption{(Color online) Partial radial distribution functions for final room-temperature states of simulated ${\rm Ni_{\alpha}Zr_{1-\alpha}}$ alloys. The cooling rate $\gamma$  is $10^{11}$ K/s. Note that panels e) ($\alpha = 0.7$) and f) ($\alpha = 0.8$) clearly reveal crystalline-like order.}
\label{fig:RDF_part_T300}
\end{figure}

\section{Results}
\subsection{Two-point correlation functions}

\begin{figure}
\centering
\includegraphics[width=0.7\columnwidth]{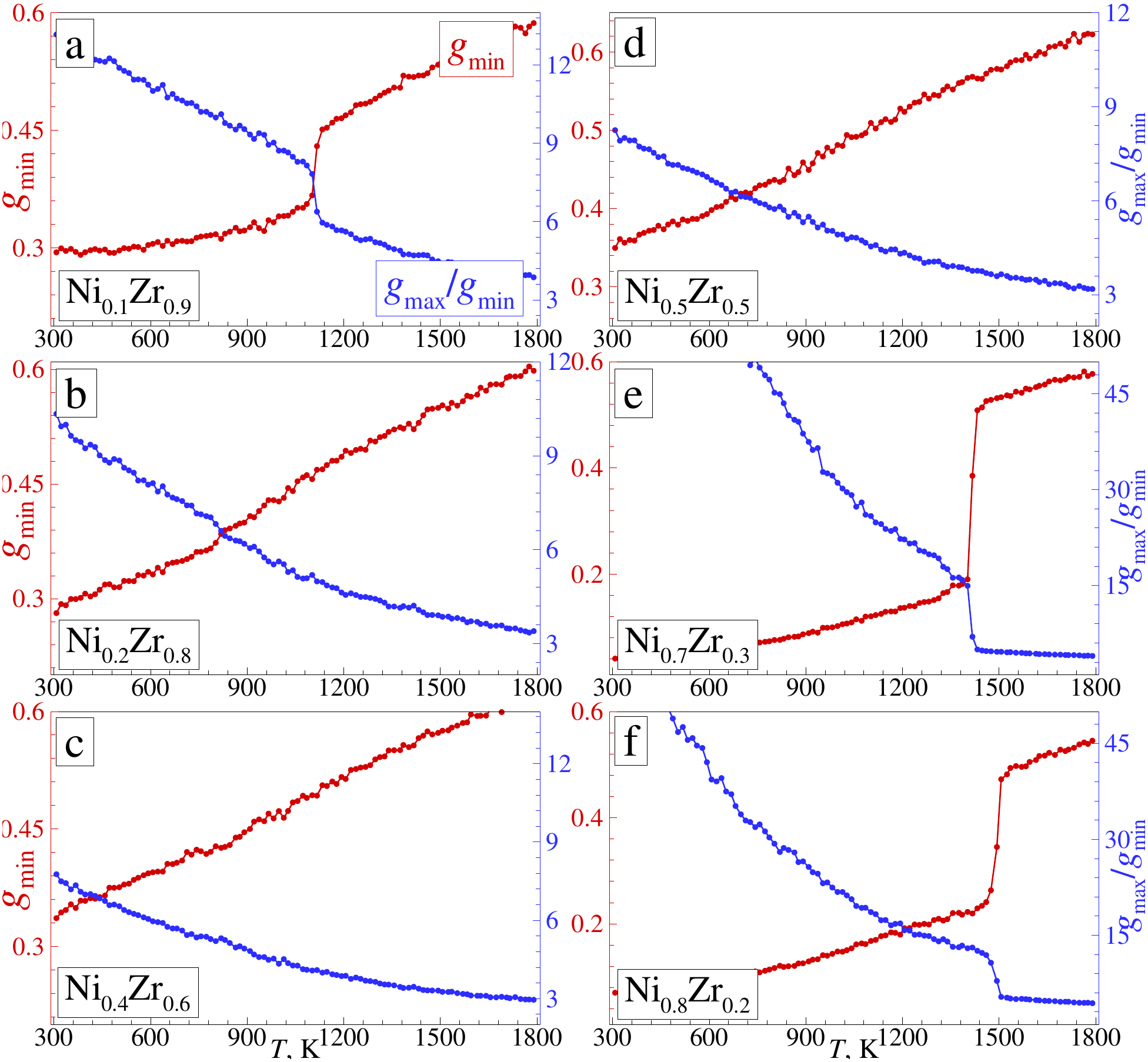}\\
\caption{(Color online) To study structural evolution of ${\rm Ni_{\alpha}Zr_{1-\alpha}}$ alloys under cooling, we show temperature dependence of structure indicators associated with the total radial distribution  function $g(r)$ (shown in Fig. 1.):  $g_{\rm min}$ - first nonzero minimum of $g(r)$ and inverted RMS freezing indicator \cite{Raveche1974JCP} $1/R = g_{\rm max}/g_{\rm min}$ ratio. Panels a) ($\alpha = 0.1$), e) ($\alpha = 0.7$) and f) ($\alpha = 0.8$) clearly reveal freezing transition to the crystalline state (the indicators sharp change in narrow temperature interval); another panels with ($0.2 \le \alpha \le 0.5$) show linear-like dependence of the indicators on $T$. Note increase of the freezing temperature with $\alpha$. The cooling rate $\gamma$  is $10^{11}$ K/s.}
\label{fig:RDF_ind}
\end{figure}

\begin{figure}
\centering
\includegraphics[width=0.8\columnwidth]{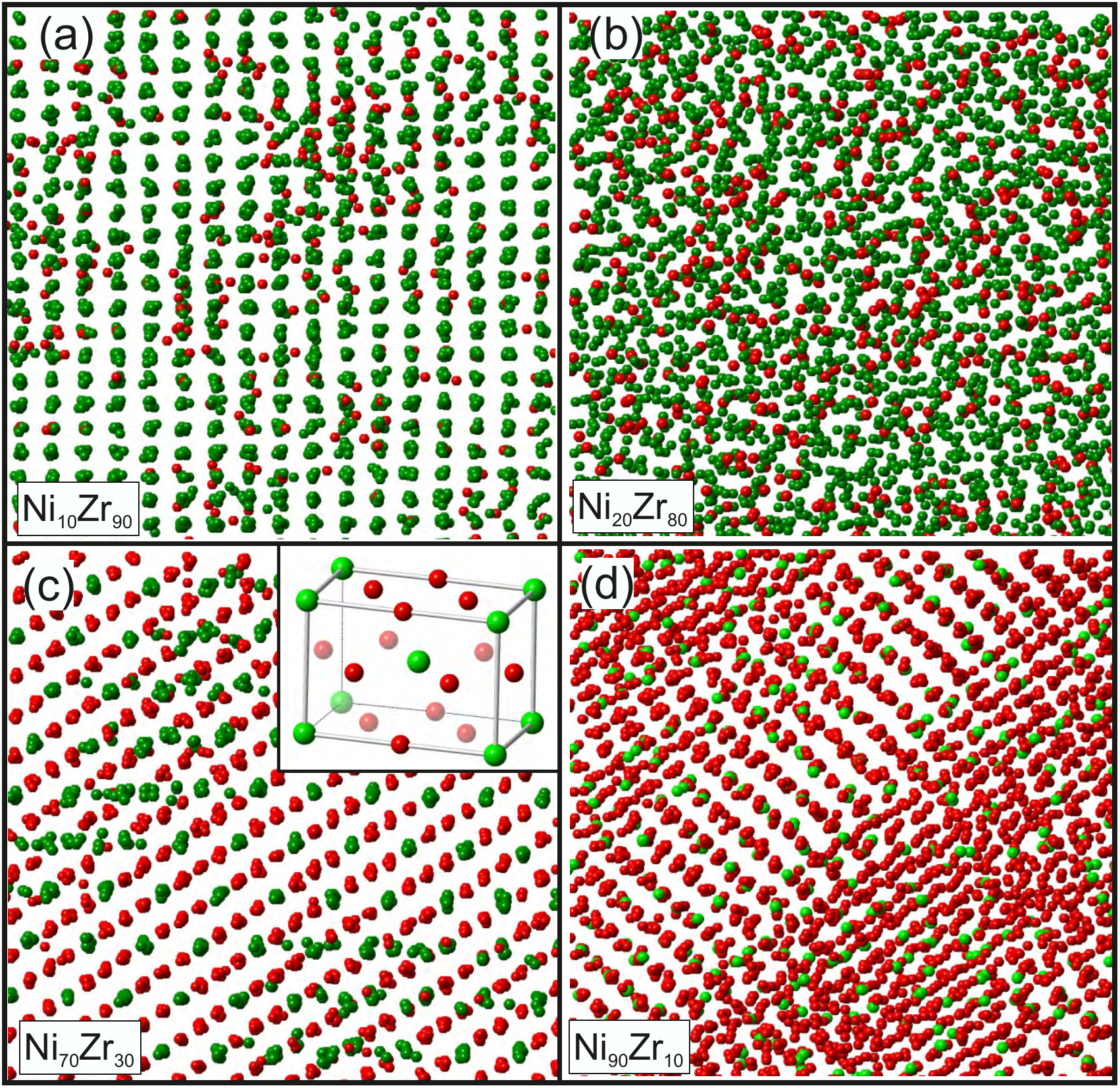}\\
\caption{(Color online) Typical snapshots of room temperature ($T=$ 300 K) states for
the simulated ${\rm Ni_{\alpha}Zr_{1-\alpha}}$ alloy quenched at cooling rate of $10^{10}$ K/s. Ni and Zr atoms are colored by red and green, respectively. (a) The structure of ${\rm Ni_{10}Zr_{90}}$ alloy is a mixture of Zr-based fcc/hcp crystal lattices with the Ni atoms mostly located in the voids; (b) ${\rm Ni_{20}Zr_{80}}$ alloy clearly is in amorphous state; (c) The structure of ${\rm Ni_{70}Zr_{30}}$  alloy corresponds to ${\rm Ni_3Zr}$ compound with tetragonal lattice (corresponding unit cell is shown in the inset);
(d) ${\rm Ni_{90}Zr_{10}}$  alloy consists of a few crystalline grains which are mixtures of fcc and hcp Ni-based lattices.}
\label{fig:snapshots}
\end{figure}

Properties of the short- and medium-range translational order can be estimated from the analysis of the total translational two-point correlation function -- radial distribution function (RDF) $g(r)$. Fig.~\ref{fig:RDF_tot_vs_T} shows temperature evolution of the total RDF of simulated ${\rm Ni_{\alpha}Zr_{1-\alpha}}$ alloys quenched at cooling rate $\gamma = 10^{11}$ K/s for a few different Ni abundances $\alpha$. Additionally, the cumulative RDFs $N(<r)=4\pi\rho\int_0^r{\xi^2g(\xi)d\xi}$ are plotted; by using $N(<r)$ it is easy to estimate the mean nearest neighbours number $N_{\rm nn}$ in the first coordination shell. All curves are color coded via temperature $T$ of the alloy (indicated on the plot). Fig.~\ref{fig:RDF_part_T300} shows both total and partial RDFs for the final room-temperature states at $T=300$ K. Shape of presented RDFs reveal disordered, glassy-like final state (taken at $T =$ 300 K) of the system for $\alpha < 0.7$ (see panels (a)-(d)) but clearly crystalline-like behaviour at $\alpha \geq 0.7$ (panels (e),(f)).
However, more rigorous analysis will show that the configuration at low $\alpha$ are also crystalline ones (see Fig.~\ref{fig:RDF_ind}, Fig.~\ref{fig:snapshots}) and Fig.~\ref{fig:phi-theta}).

To identify structural phase transitions in a system by using the total RDF shape, it is convenient to use inverted Ravech\'{e}-Mountain-Streett (RMS) parameter \cite{Raveche1974JCP} $1/R$ which is defined as ratio of $g(r)$ value at the first maximum to that at the first nonzero minimum: $1/R = g_{\rm max}/g_{\rm min}$. However, the use of the parameter $g_{\rm min}$ instead may be even more convenient because it is nearly independent on the interaction potential softness (unlike the parameter RMS) \cite{Khrapak2017SciRep}. These parameters were recently used to identify the liquid-glass transition in CuZr system \cite{Ryltsev2016JCP}. Fig.~\ref{fig:RDF_ind} shows how both $R^{-1}$ and $g_{\rm min}$ vary at cooling of the Ni-Zr alloys (at the same $\alpha$ values as in Fig.~\ref{fig:RDF_tot_vs_T}). Both indicators clearly show transition to crystalline structure at cooling for $\alpha=0.1, 0.7, 0.9$ (panels (a), (e), (f)) but continuous evolution of the structure to the disordered amorphous solid state for $ 0.15 < \alpha < 0.66$ (panels b,c,d). Note that temperature dependencies of the structure indicators shown in Fig.~\ref{fig:RDF_ind}b,c,d do not demonstrate any kinks or drastic inflections indicating glass transition as for Cu-Zr alloys (see, for example \cite{Ryltsev2016JCP}). The possible reason is the lack of pronounced icosahedral ordering in Ni-Zr alloys (see below) whose increase at the glass transition temperature leads to the mentioned anomalies in the case of Cu-Zr.

\begin{figure}
  \centering
  \includegraphics[width=0.7\columnwidth]{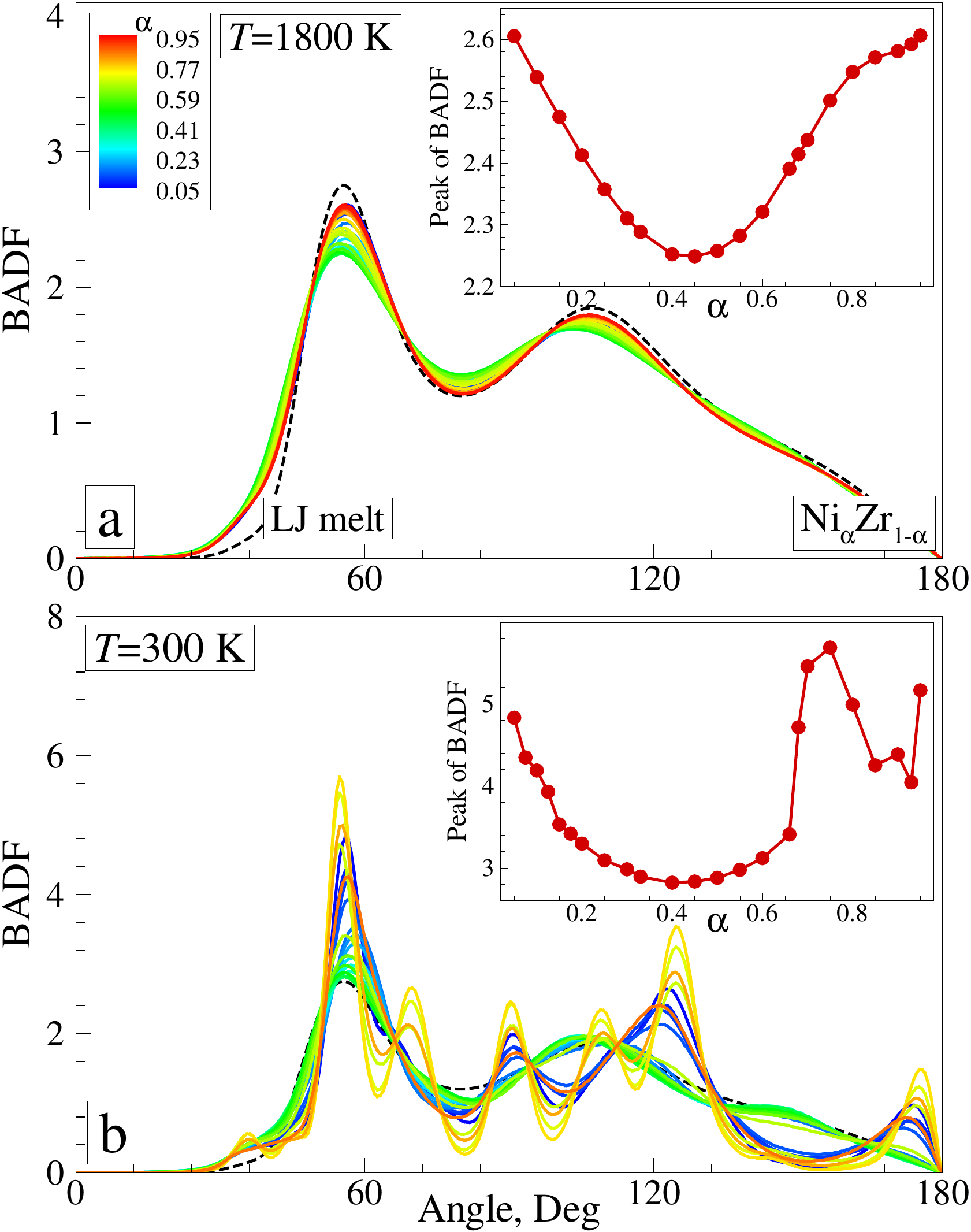}\\
  \caption{(Color online)  Bond angle distribution functions (BADF) of simulated ${\rm Ni_{\alpha}Zr_{1-\alpha}}$ alloys calculated at different temperatures $T$: (a) melts at 1800 K and solid phases taken at 300 K (b).
  The curves are color-coded via $\alpha$ value. Black dashed line shows BADF for the Lennard-Jones melt (which is nearly universal along the melting curve).  Insets show dependence of the BADF peak on the $\alpha$.
  The cooling rate $\gamma$  is $10^{11}$ K/s.}
  \label{fig:BADF}
\end{figure}

\begin{figure*}
  \centering
  \includegraphics[width=0.3\textwidth]{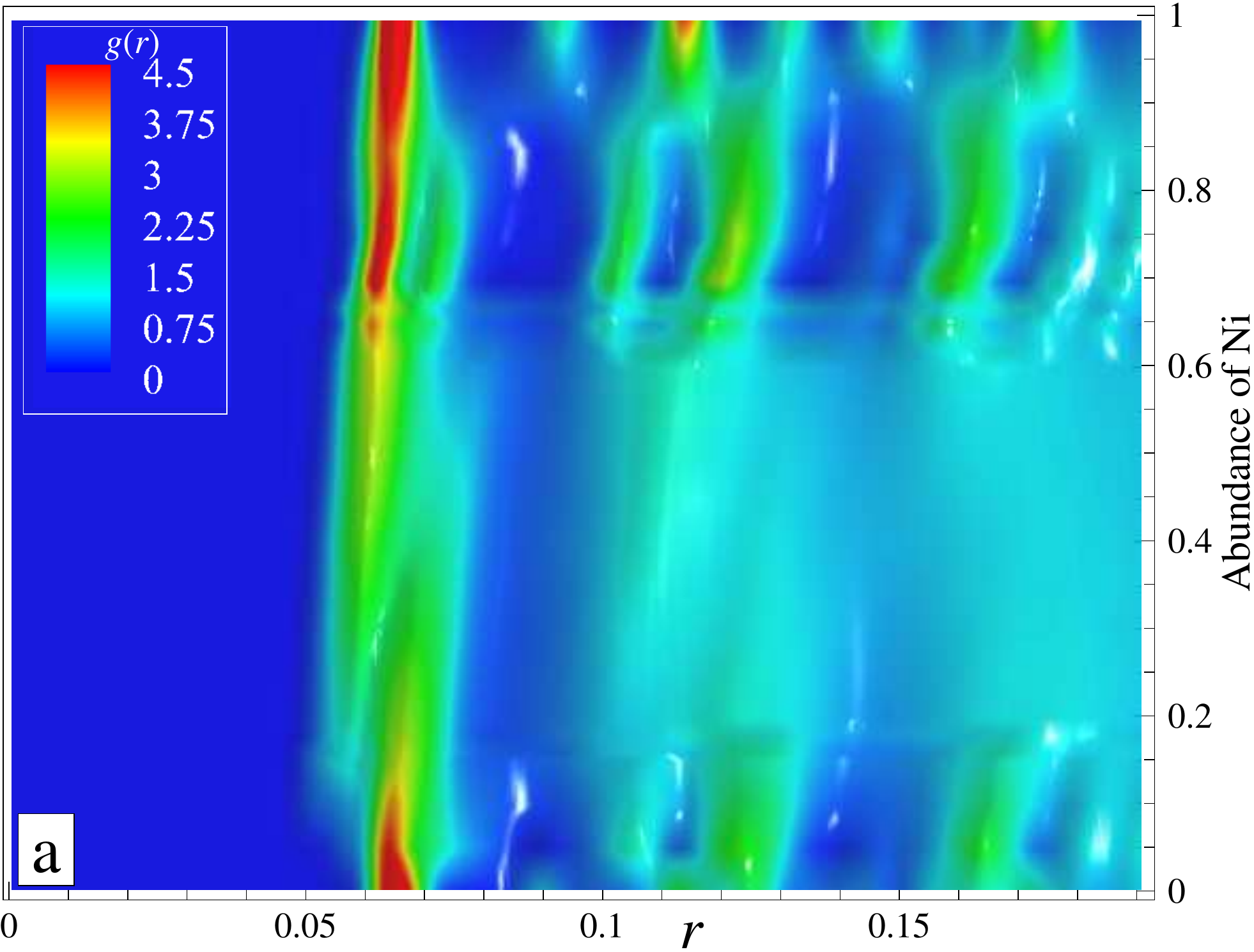}
  \includegraphics[width=0.3\textwidth]{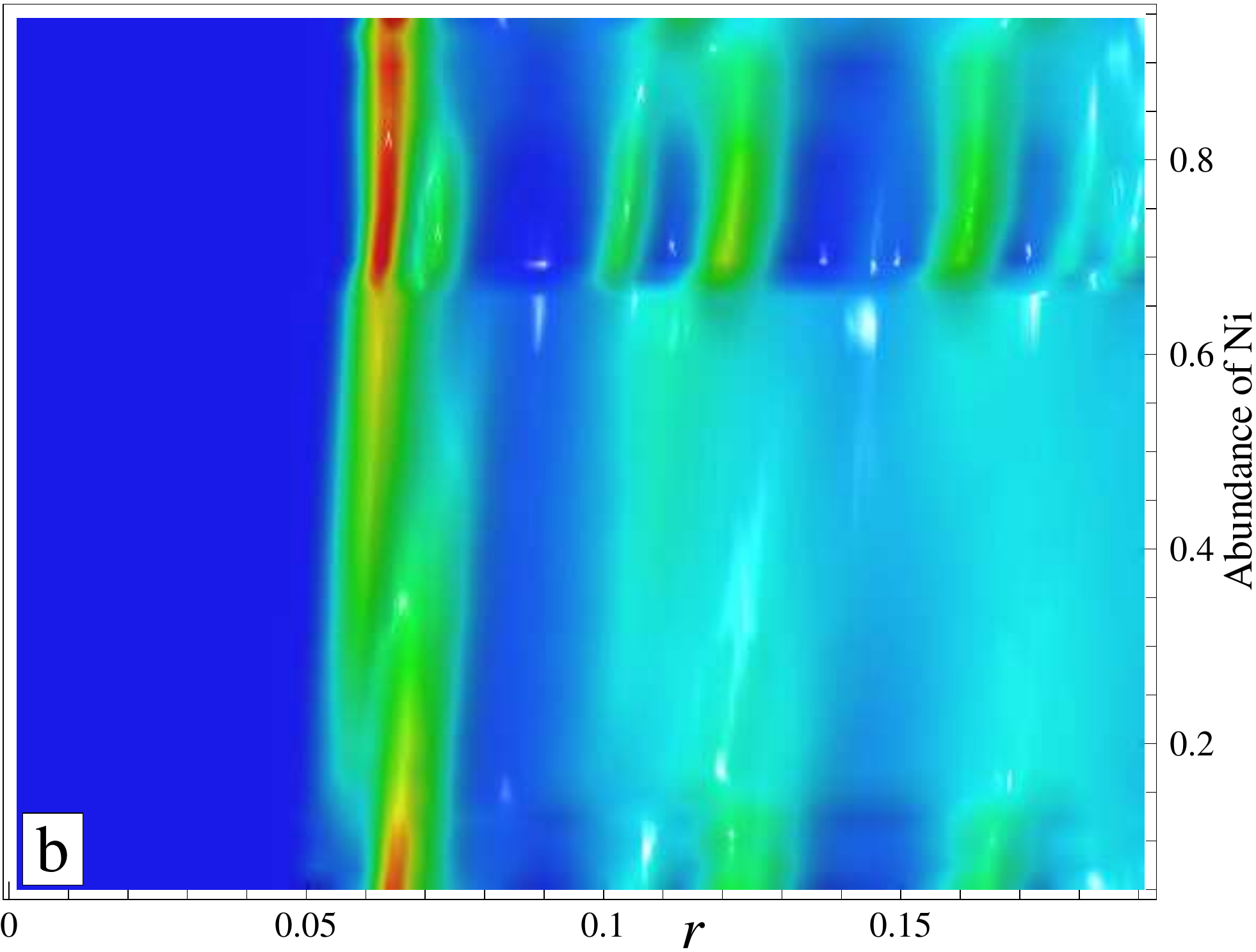}
   \includegraphics[width=0.3\textwidth]{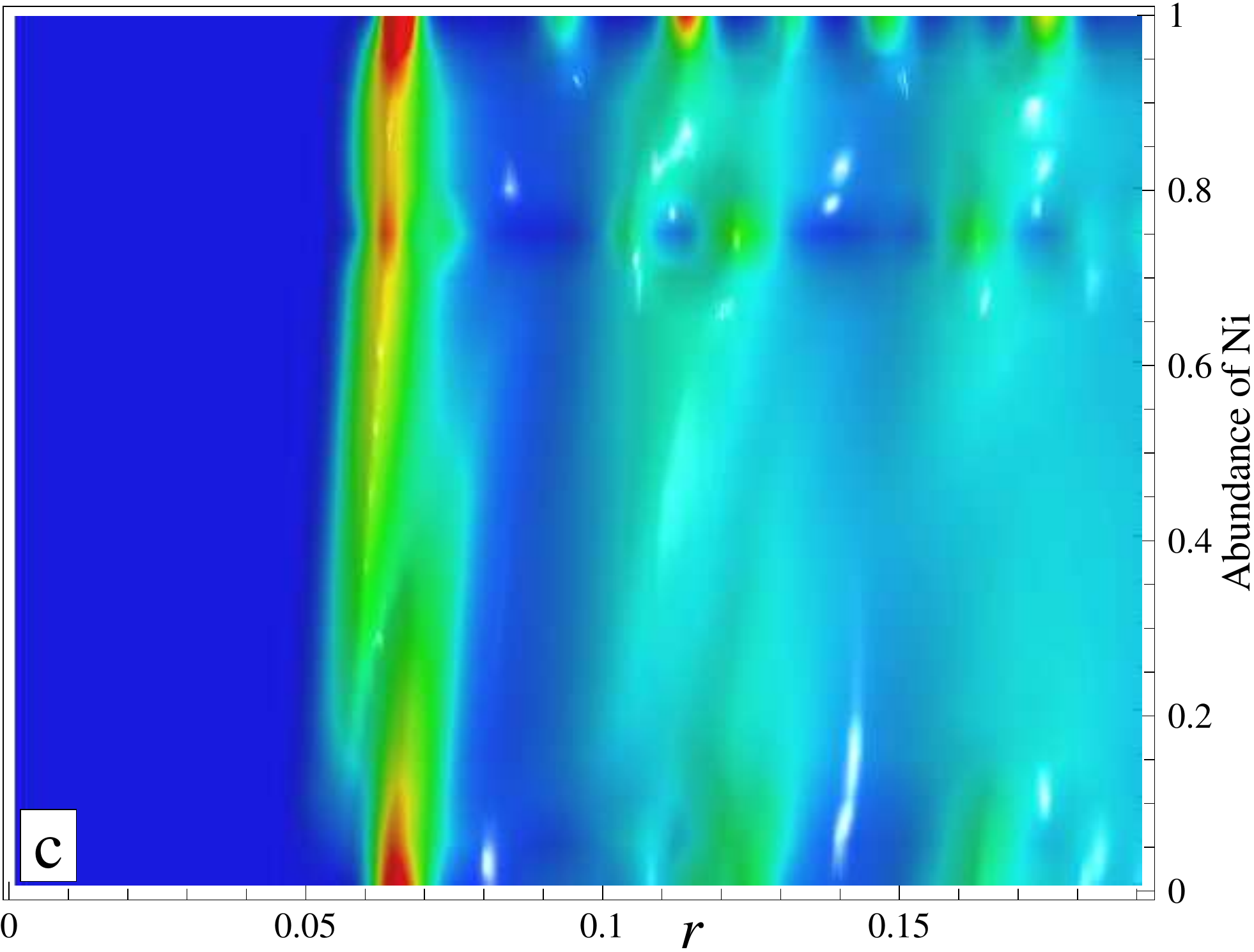}
   \includegraphics[width=0.3\textwidth]{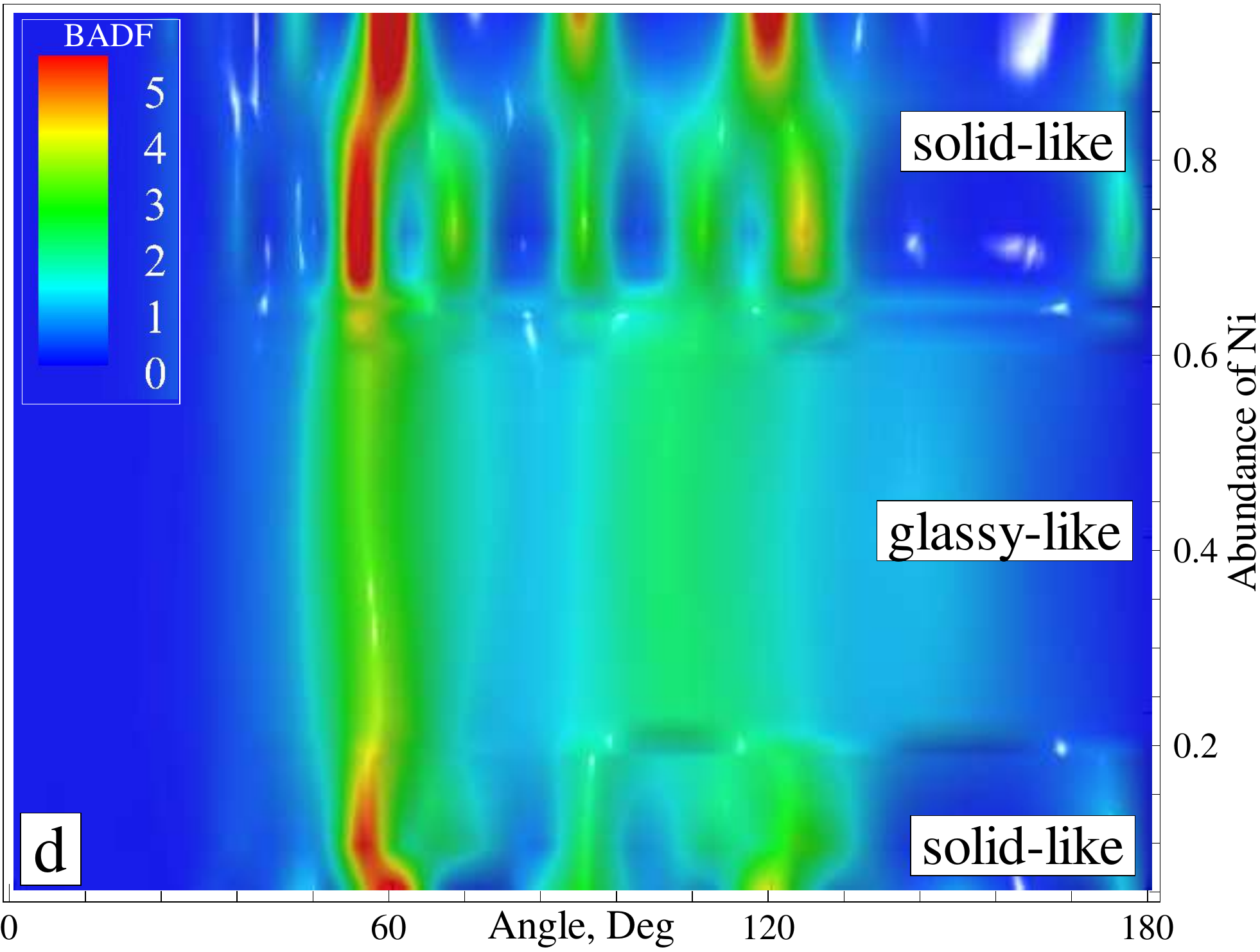}
  \includegraphics[width=0.3\textwidth]{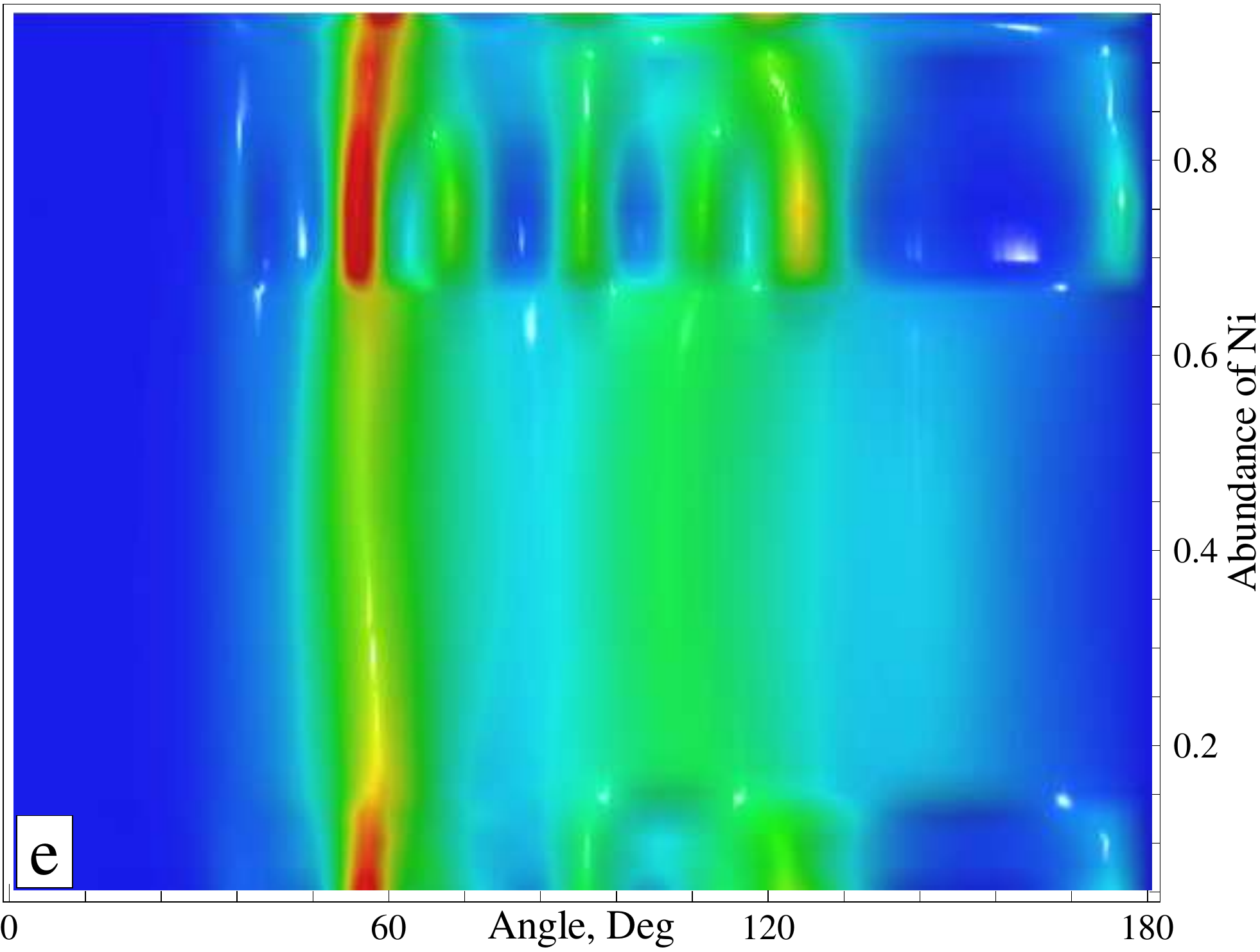}
   \includegraphics[width=0.3\textwidth]{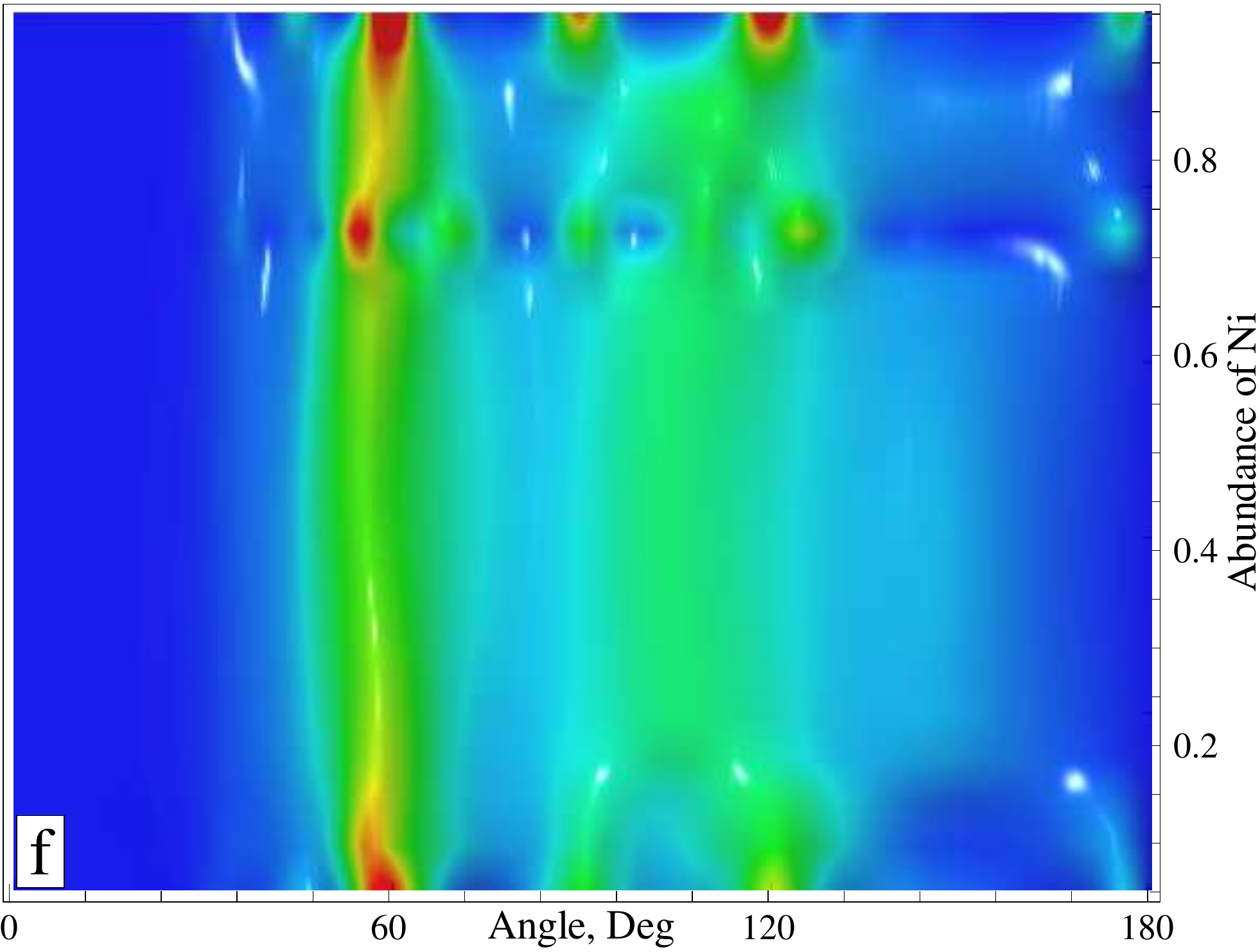}
  \caption{(Color online) Both radial distribution functions $g(r)$ (top panels) and bond angle distribution functions $P(\phi)$  (bottom panels) for  ${\rm Ni_{\alpha}Zr_{1-\alpha}}$ alloys at room temperature $T=$ 300 K plotted
  versus Ni abundance $\alpha$. The cooling rates $\gamma$ (from left to right) are $10^{10}$, $10^{11}$ and $10^{12}$ K/s. The color on the planes represents the $g(r)$ and $P(\phi)$ values (see the scales in panels (a) and (d), respectively).}
  \label{fig:color_map}
\end{figure*}

Another important two-point correlation function, bond angle distribution function (BADF) $P(\phi)$, measures the probability that two nearest neighbours and central atom form the angle $\phi$. Fig.~\ref{fig:BADF} shows $P(\phi)$ at different Ni compositions $\alpha$ for both the initial ($T=1800$ K) and final ($T=300$ K) states. The room-temperature BADFs demonstrate the similar picture as RDFs: there are two crystal-like concentration regions and the glassy one in between them.

Fig.~\ref{fig:BADF} (a) allows analyzing the structure of high-temperature states from which the system was cooled down. Comparing BADF for Ni-Zr system with that for Lennard-Jones melt at $T^* \approx 1.5$, $\rho^* \approx 1$ (in reduced Lennard-Jones units \cite{Ryltsev2013PRE}) we see their close similarity. That suggests the idea that BADFs of different close-packed systems are very similar for the melts near the solid-liquid coexistence line. Such similarity has been already observed for Cu-Zr alloys \cite{Ryltsev2016JCP,Klumov2016JETPLett}. We argue that such similarity can be explained by the existence of local tetrahedral order which has been found in Lennard-Jones fluids \cite{Ryltsev2013PRE}. Our results suggest the formation of local tetrahedra might be an universal feature of close-packed fluids.

The insets in Fig.~\ref{fig:BADF} show how the first maximum value of the BADF $P^{\rm m}(\alpha)$ depends on the Ni abundance $\alpha$. It is a new measure characterizing the mean local order of the system. At high temperature (panel (a)), there is a minimum of  $P^{\rm m}(\alpha)$ at $\alpha\approx 0.45$ ($P^{\rm m} \approx 2.25$); at room temperature (panel (b)), the minimum occurs at the same $\alpha$ with $P^{\rm m} \approx 2.8$ which is very close to that value for the Lennard-Jones melt ($P^{\rm m}_{\rm LJ} \approx 2.75$). So $P^{\rm m}$ value looks as very promising measure to quantify the structural order in the system. For instance, $P^{\rm m}(\alpha)$ dependence clearly shows inflections at $\alpha\approx 0.15$, $\alpha\approx 0.65$ corresponding to boundaries between crystalline and glassy regions for room-temperature states (see inset in Fig.~\ref{fig:BADF}b).

Thus, the analysis of two-point correlation functions allows estimating the composition range corresponding to glass formation and analysing the structure of crystal phases.  To demonstrate cooling rate dependence of structure of quenched solid phases, we show in Fig.~\ref{fig:color_map} color-coded plots of both the RDFs and BADFs for atomic configurations obtained at three different cooling rates and different concentrations. We see that the concentration range corresponding to glassy state essentially depends on cooling rate. Indeed it is about $\alpha\in (0.2, 0.65)$ for $10^{10}$ K/s but $\alpha\in (0.1, 0.95)$ for $10^{12}$ K/s. At $\gamma=10^{12}$ K/s and $\alpha \approx 0.75$ the crystal state ''island'' included to the glassy state area can be seen. That means the intermetallic compound ${\rm Ni_3Zr}$ has relatively high critical formation rate. Another interesting feature is observed in the area corresponding to crystal states: at high Ni concentration the change of crystal lattice structure takes place. This feature is more pronounced at the lowest cooling rate $\gamma=10^{10}$ K/s (Fig.~\ref{fig:color_map} (a) and (d)). Also note that RDFs obtained at all cooling rates studied demonstrate the change of the first peak locations at $\alpha \approx 0.45-0.5$. It is interesting that the minimum on concentration dependence of first maximum value of BADF is observed at the same composition (see insert in Fig.~\ref{fig:BADF} (a)).

The results revealed from two-point correlation functions are confirmed by visual analysis of simulated configurations. In Fig.~\ref{fig:snapshots} we show typical snapshots for final room-temperature states obtained at $\gamma=10^{10}$ K/s and different compositions. Each picture represents typical structure for characteristic concentration range revealed by  two-point correlation functions analysis.  The structure of Zr-rich alloys is a mixture of Zr-based fcc/hcp crystal lattices (see also the Fig.~\ref{fig:q4-q6_T300}) with the Ni atoms mostly located in the voids (see Fig.~\ref{fig:snapshots}a for $\textrm{Ni}_{10}\textrm{Zr}_{90}$ alloy). The alloys at $ 0.2 <\alpha < 0.65$ (at $\gamma = 10^{10}$ K/s) are visually amorphous (see Fig.~\ref{fig:snapshots}b). The structure of alloys at $\alpha \simeq 0.75$ corresponds to $\textrm{Ni}_{3}\textrm{Zr}$ compound with tetragonal lattice (see the unit cell in the inset of Fig.~\ref{fig:snapshots}c). This structure has I4/mmm space group and, for example, corresponds to $\textrm{Li}_{3}\textrm{(Al,Be)}$ compounds. A number of Zr atoms which do not match $\textrm{Ni}_{3}\textrm{Zr}$ stoichiometry are located in voids (Fig.~\ref{fig:snapshots}c). The structure of Ni-rich alloys contains a few crystalline grains (see Fig.~\ref{fig:snapshots}d for $\textrm{Ni}_{90}\textrm{Zr}_{10}$  alloy).  Each grain is a mixture of fcc and hcp Ni-based lattices (as well as for Zr-rich alloys). However, unlike Zr-rich alloys, the atoms of second component (Zr) incorporate into the solvent crystal lattice substitutionally, by replacing Ni particles in the lattice.

\subsection{Local orientational order}
\begin{figure}
\centering
\includegraphics[width=0.7\columnwidth]{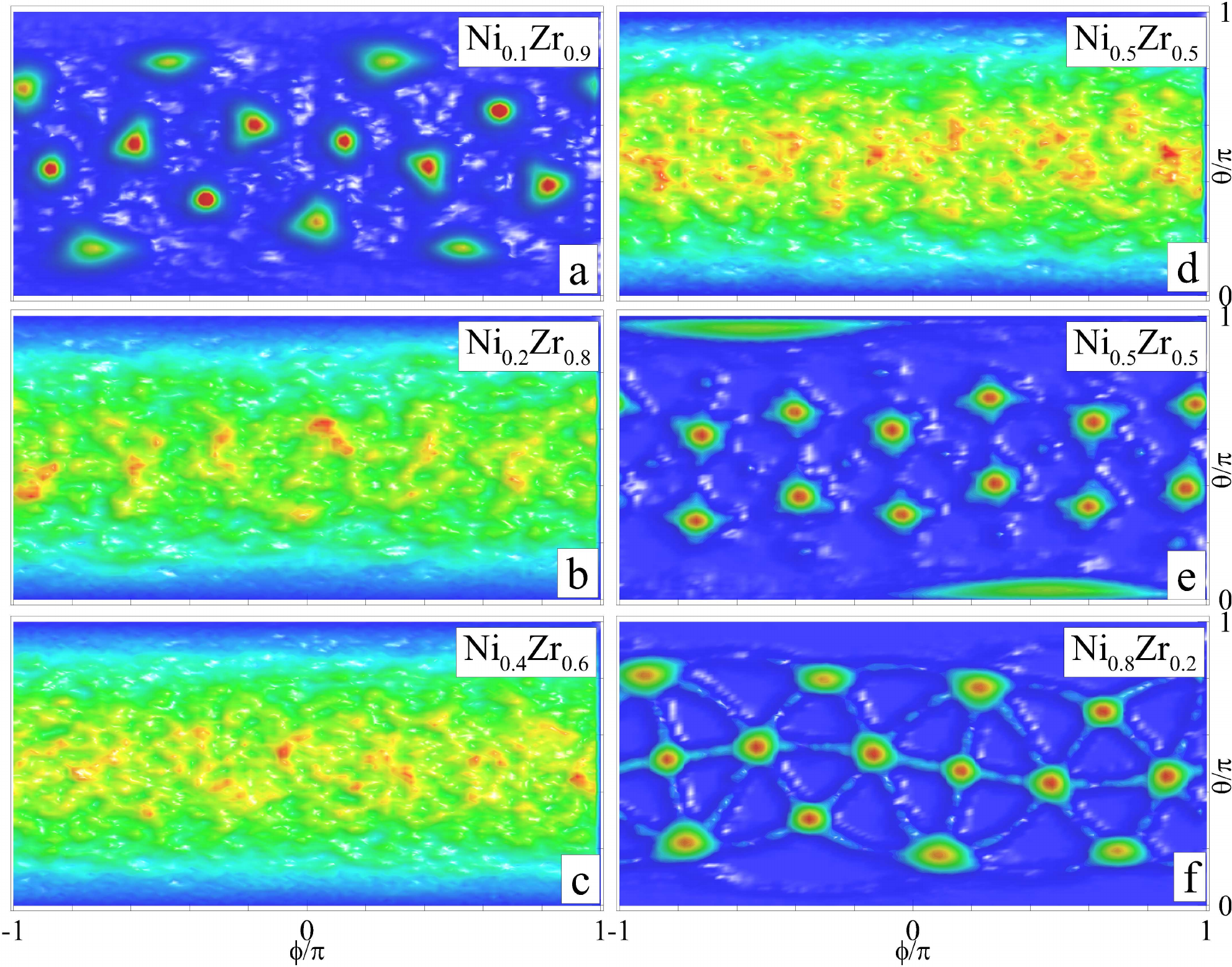}\\
\caption{(Color online) In order to show structural difference between ${\rm Ni_{\alpha}Zr_{1-\alpha}}$ alloys at different Ni abundances $\alpha$, we show probability distributions of 12 nearest neighbours
on the ($\phi-\theta$) plane, (where $\phi,~\theta$ - are the polar and azimuthal angles of the neighboring atoms, respectively). The probabilities are shown for final room-temperature states of the alloys quenched at cooling rate of $10^{11}$ K/s. Panels (b), (c), (d) with $0.2 \le \alpha \le 0.5$ clearly reveal disordered glassy-like structure while panels (a) ($\alpha = 0.1$), (e) ($\alpha = 0.7$), (f) ($\alpha = 0.8$) reveal the crystalline-like local order.}
\label{fig:phi-theta}
\end{figure}

Two-point correlation functions analyzed above do not allow studying fine details of orientational order. For that purpose, we use the BOOP method described in section \ref{sec_methods}. Calculation of rotation invariants of both second $q_l$ and third $w_l$ order requires specifying the number of nearest neighbours $N_{\rm nn}$. That number can be estimated from the behaviour of the cumulative RDF $N(<r)$ presented in Fig.~\ref{fig:RDF_tot_vs_T}. We see from the picture that $N_{\rm nn}\simeq 12$ as it typically is for close-packed systems.

 Before doing BOOP analysis, it is interesting to know how the nearest neighbors are distributed in the angle ($\phi$-$\theta$)-plane, where $\phi$ and $\theta$ are respectively the polar and the azimuthal angle (spherical coordinates) of a neighboring atom with respect to the central one. Such distributions are shown in Fig.~\ref{fig:phi-theta} for the final room-temperature states of Ni-Zr alloys at different Ni concentration $\alpha$. Panels (b), (c), (d) with $0.2 \leq \alpha \leq 0.5$ clearly reveal disordered glassy state of the alloys, while panels (a) ($\alpha = 0.1$), (e) ($\alpha = 0.7$) and (f) ($\alpha = 0.8$) demonstrate crystalline-like local order. These results are in close agreement with the previous results obtained for two-point correlation functions. Note that $\phi-\theta$ distribution is much more sensitive than $g(r)$ for distinguishing crystalline and glassy states. Indeed, the difference in $g(r)$ between partially crystalline $\textrm{Ni}_{10}\textrm{Zr}_{90}$ alloy (Fig.~\ref{fig:RDF_ind}a) and completely glassy $\textrm{Ni}_{20}\textrm{Zr}_{80}$ one (Fig.~\ref{fig:RDF_ind}b) is not obvious whereas this difference is clearly seen from $\phi-\theta$ distribution (compare Fig.~\ref{fig:phi-theta}a and Fig.~\ref{fig:phi-theta}b).

\begin{figure}
\centering
\includegraphics[width=0.4\textwidth]{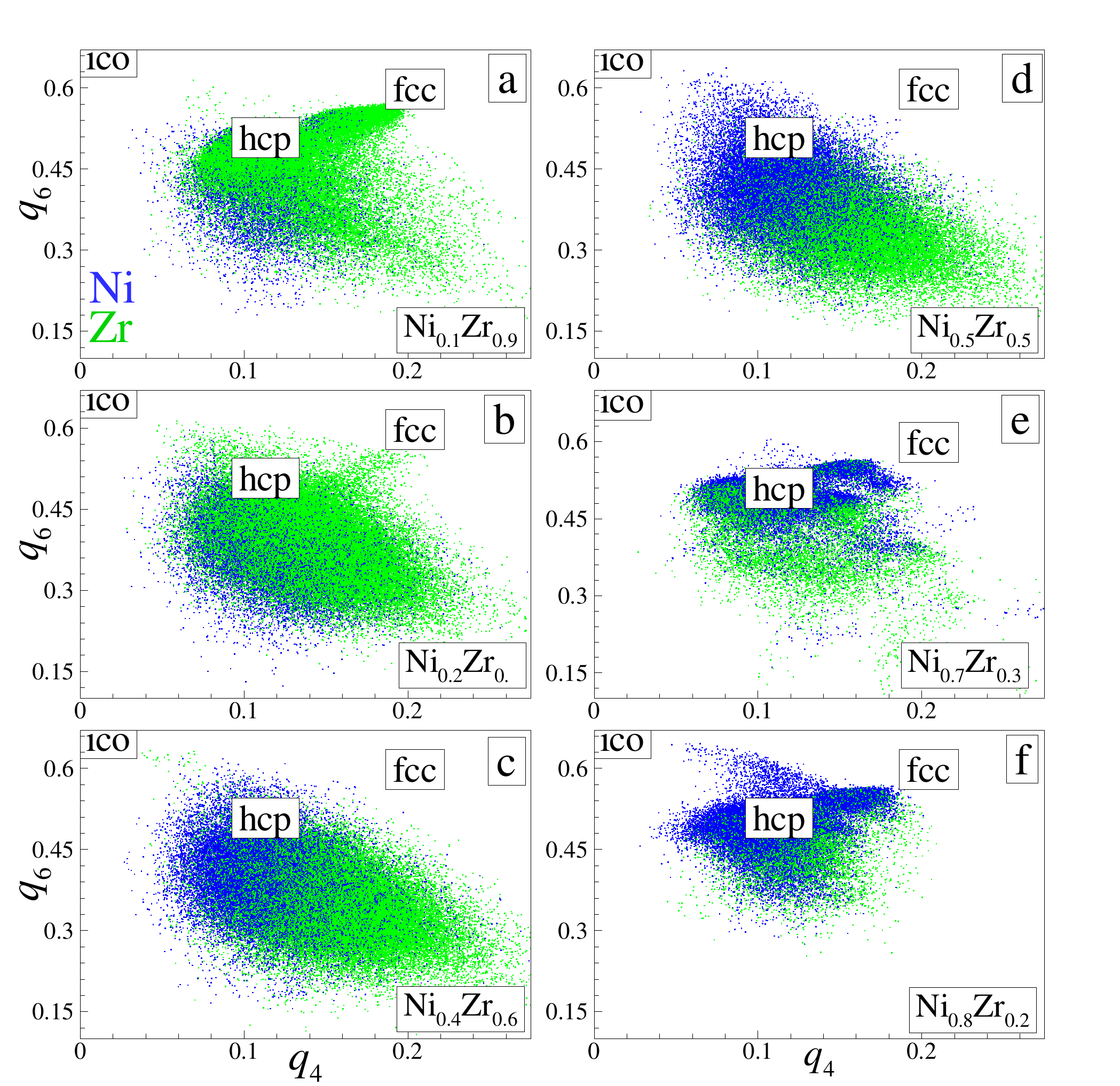}\\
\caption{(Color online). Local orientational order of the ${\rm Ni_{\alpha}Zr_{1-\alpha}}$ alloys on the $q_4\--q_6$ plane at different Ni abundances $\alpha$. Bond orientational order parameters (BOOP) were calculated via 12 nearest neighbours for both nickel-centered (blue) and zirconium-centered (green) atoms to identify different close packed structures. BOOP for the perfect icosahedron, hcp and fcc clusters are also indicated for the comparison. Temperature of the system is $T=300$ K. The cooling rate is $\gamma$ is $10^{11}$ K/s.}
\label{fig:q4-q6_T300}
\end{figure}

\begin{figure}
\centering
\includegraphics[width=0.4\textwidth]{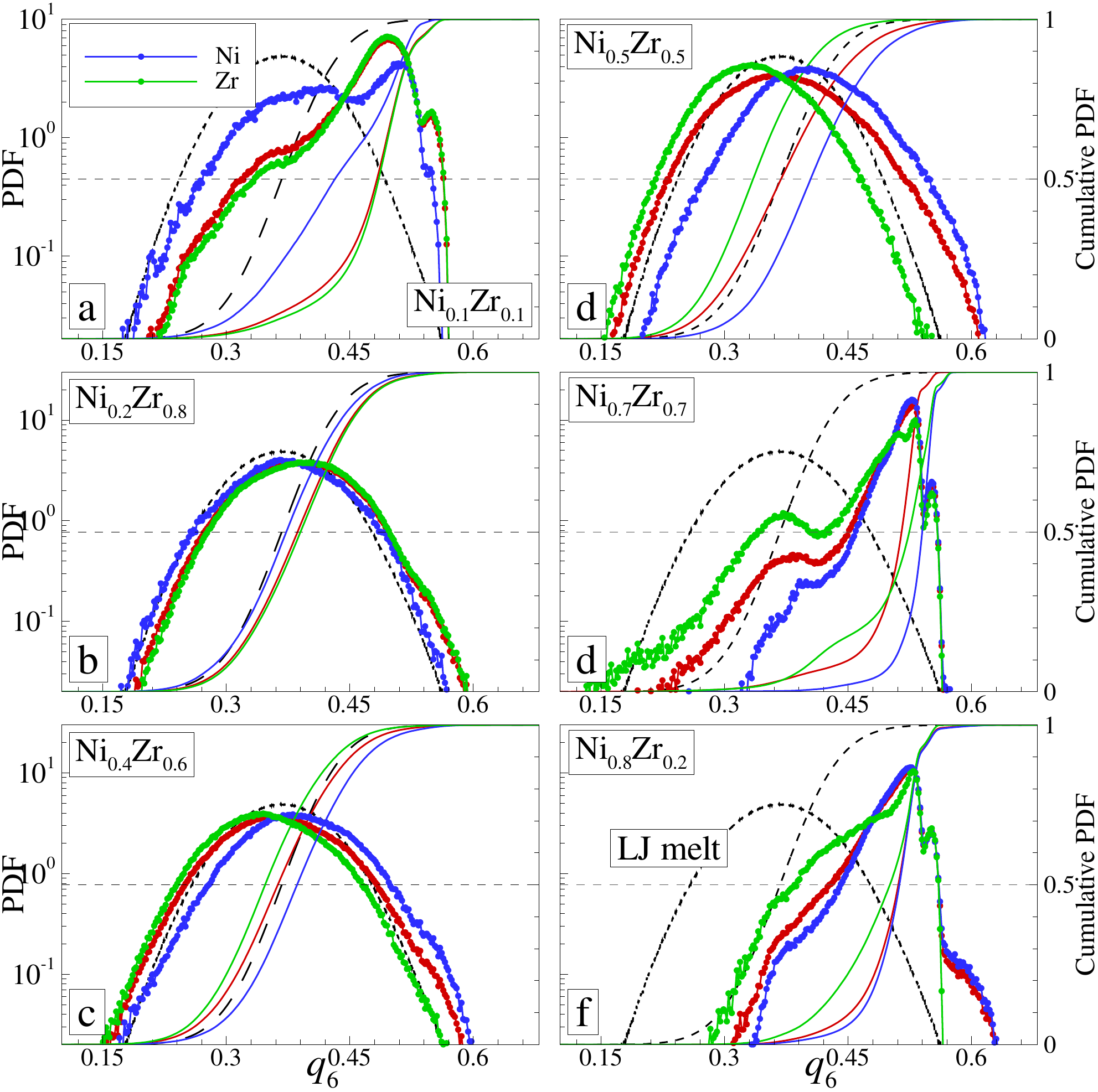}\\
\caption{(Color online). Local orientational order of the quenched ${\rm Ni_{\alpha}Zr_{1-\alpha}}$ alloy taken at T=300 K. Normalized probability distributions functions $P(q_6)$ (PDFs) for nickel-centered (blue color), zirconium-centered (green color) atoms and total $P(q_6)$ (red color) are plotted for different $\alpha$ values (indicated on the plot). Cumulative PDFs are also plotted to quantify the phase state of the system. Orientational order parameter $q_6$ was calculated by using 12 nearest neighbours. Additionally, the same distributions obtained for the Lennard-Jones (LJ) melt (the PDFs are nearly universal along the LJ melting curve) are plotted on each panel (black dashed curves) for the comparison. The cooling rate $\gamma$ is $10^{11}$ K/s).}
\label{fig:pdf_q6}
\end{figure}

\begin{figure}
\centering
\includegraphics[width=0.4\textwidth]{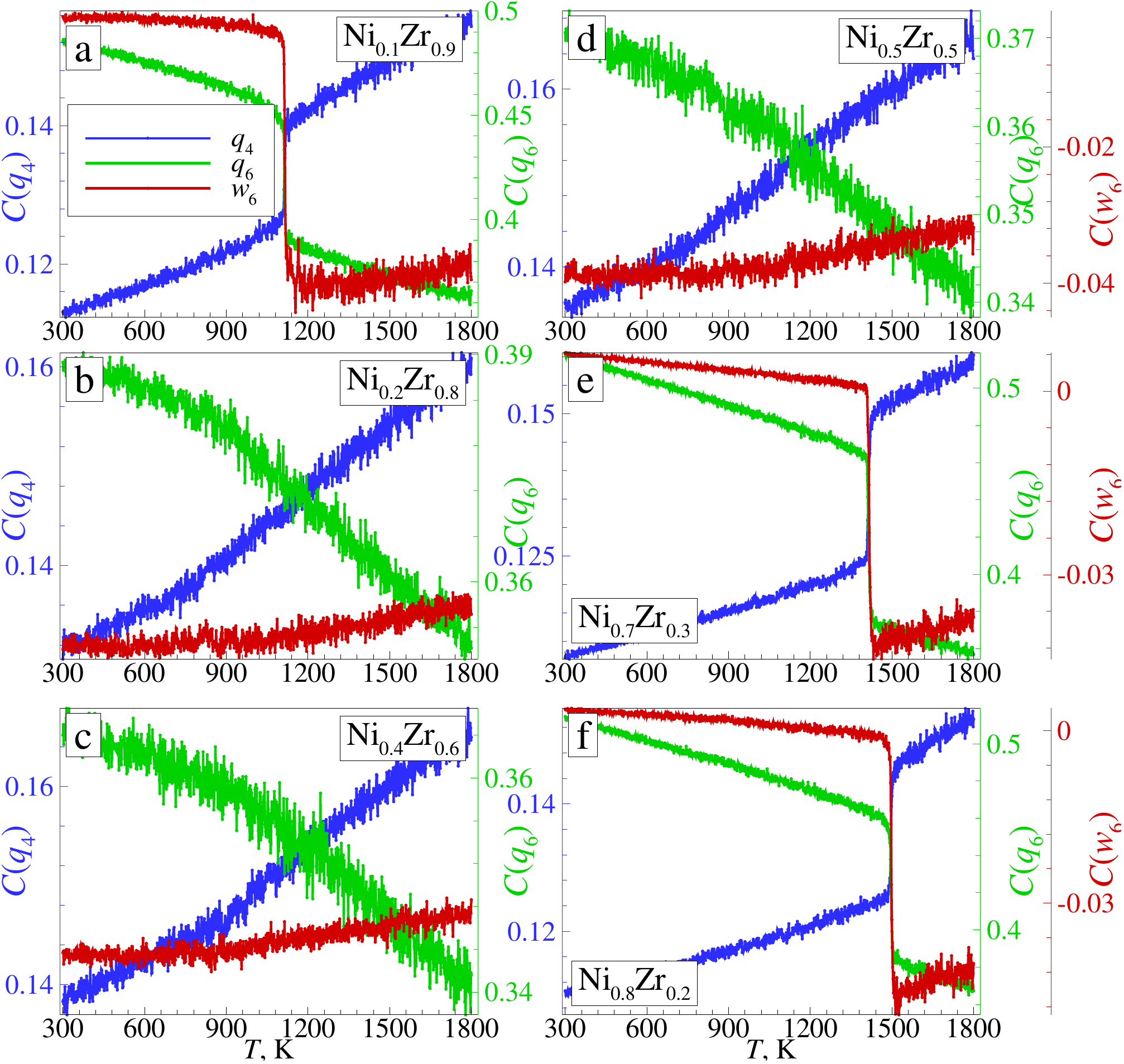}\\
\caption{(Color online). Cumulants of both $P(q_l)$ and $P(w_l)$ distributions for ${\rm Ni_{\alpha}Zr_{1-\alpha}}$ alloys plotted versus temperature $T$ at different nickel abundances $\alpha$.
Sharp changes of the presented cumulants ($q_4$, $q_6$, $w_6$) observed in the panels (a, e, f) reveal freezing transition.  Note the significant increase of the crystallization temperature with  an increase of $\alpha$ value. Another panels reveal liquid-glass transition under cooling. The cooling rate $\gamma$ is $10^{11}$ K/s.}
\label{fig:cums}
\end{figure}

\begin{figure}
  \centering
  \includegraphics[width=0.75\columnwidth]{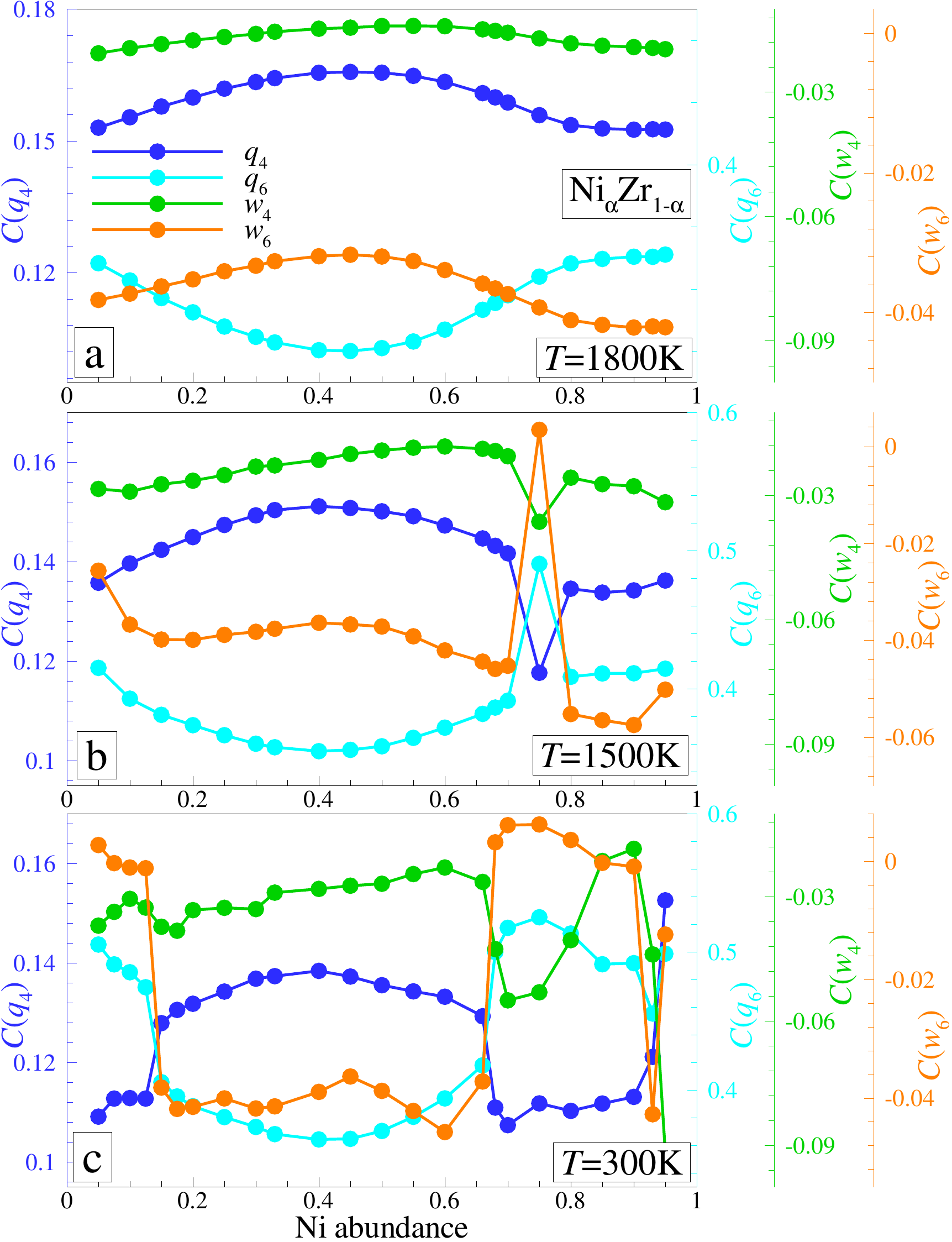}\\
  \caption{(Color online) Cumulants of both $P(q_l)$ and $P(w_l)$ distributions for ${\rm Ni_{\alpha}Zr_{1-\alpha}}$ alloys plotted versus Ni abundance $\alpha$ at different temperatures $T$ (indicated on the plot). The cooling rate $\gamma$ is $10^{11}$ K/s.}
  \label{fig:cum}
\end{figure}

Fig.~\ref{fig:q4-q6_T300} shows local orientational order of the final room-temperature states of simulated ${\rm Cu_{\alpha}Zr_{1-\alpha}}$ alloys (obtained at $\gamma=10^{11}$ K/s) on the plane of rotational invariants $q_4\--q_6$ (calculated via fixed number of nearest neighbors: $N_{\rm nn}$ = 12) at few $\alpha$ values. Points in the figure correspond to $(q_4, q_6)$ values calculated for the nearest neighbors of each atom. Positions of $(q_4, q_6)$ for the perfect hcp, fcc and icosahedral clusters are also marked. The picture clearly shows the formation of small amount of both hcp and fcc structures at low Ni concentration (Fig.~\ref{fig:q4-q6_T300} (a)), mostly disordered liquid-like behavior (with traces of hcp-like and fcc-like clusters) at intermediate $\alpha$ (Fig.~\ref{fig:q4-q6_T300} (b),(c)) and strong ordering (with different crystalline symmetries) at high $\alpha$ values (Fig.~\ref{fig:q4-q6_T300} (d)). Some traces of distorted icosahedral particles are present in Fig.~\ref{fig:q4-q6_T300} (panel (e)) at $\alpha=0.8$.

\begin{figure}
\centering
\includegraphics[width=0.4\textwidth]{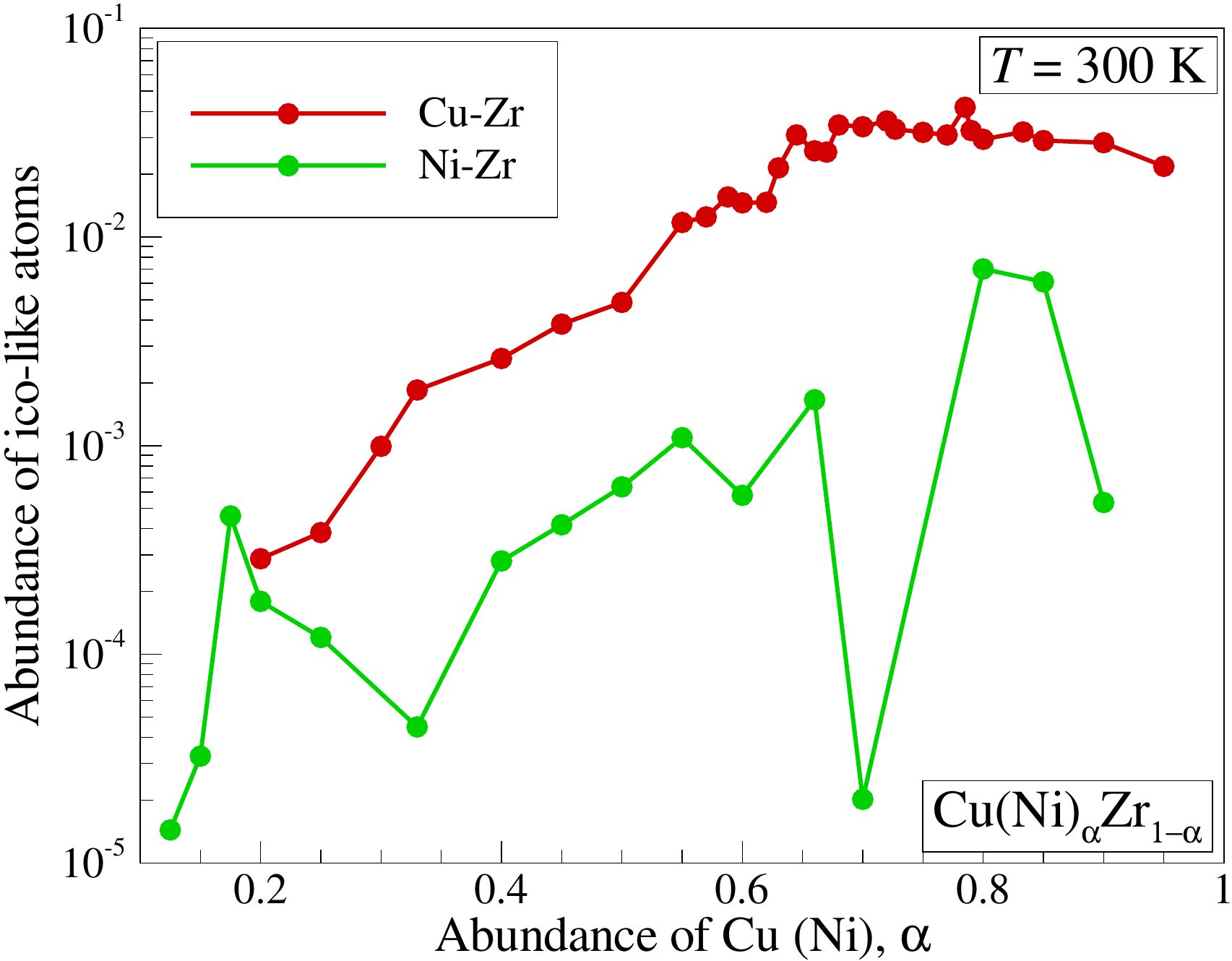}\\
\caption{(Color online). Abundance of icosahedral-like clusters $n_{\rm ico}$ plotted versus copper (nickel) concentration $\alpha$ for simulated Cu(Ni)-Zr alloys at $T=300$ K. Cooling rate is $10^{11}$ K/s for both system.}
\label{fig:ico_vs_x}
\end{figure}

To quantify the distributions in Fig.~\ref{fig:q4-q6_T300}, we use the normalized one-dimensional probability density function over different rotational invariants $P(q_l)$ and $P(w_l)$, so that (for e.g. $q_l$) we have $\int_{-\infty}^{\infty} P(q_l)dq_l \equiv 1$. Even more convenient is using (cumulative) distribution functions associated with the $P(q_l)$ and $P(w_l)$~\cite{KlumovPU10,KlumovPRB11} which are defined as (by using as example the bond order parameter $q_6$): $C_q(x)=\int_{-\infty}^x P(q_6)dq_6$. Using the cumulative functions we can find abundance (density) of any structure with given accuracy $\delta_{\rm cr}$: $C_q(q^{\rm cr}+\delta_{\rm cr})$ - $C_q(q^{\rm cr}-\delta_{\rm cr})$, where $q^{\rm cr}$ is the bond order parameter for an ideal structure. We note, that the set of distributions $P$ and $C$ taken for different $q_l$ and $w_l$ completely describes the local orientational order in the system as an abundance of different ordered and disordered structures. Fig.~\ref{fig:pdf_q6} shows the probability distributions $P(q_6)$ of the quenched Ni-Zr system (with cooling rate $\gamma=10^{11}$ K/s) taken at room temperature and its cumulative distributions $C(q_6)$ at different $\alpha$ values. Corresponding distributions for the Lennard-Jones melt are also plotted for the comparison. Again, these distributions clearly show disordered states within the range of $\alpha \in (0.2,0.6)$ and crystalline-like states outside this range. Interestingly, the disordered $P(q_6)$ distributions for the Ni-Zr alloys are very close to those for the Lennard-Jones melt. {It suggests that local orientational order of LJ melt (which is nearly universal along the melting curve) can be used to understand local order in amorphous solids too.}

Evidently, the cumulative distribution $C^l_q(x)$ is the abundance of atoms, having $q_l<x$ and $C^l_q(\infty)\equiv 1$. Cumulants of these distributions define the global order parameters. E.g. cumulant $C(q_6)$ is the position of the half-height of the cumulative distribution $C^6_q(x)$, so that $C^6_q(C(q6)) \equiv 1/2$. It was shown that
such cumulants calculated for different invariants $q_l$ and $w_l$ are very sensitive to structural transitions \cite{KlumovPU10}. Fig.~\ref{fig:cums} shows temperature dependencies of these cumulants obtained at cooling for different values of $\alpha$. Again, panels (a), (e) and (f) clearly show transition to crystalline state; another panels reveal amorphization of the alloys under consideration. Note that, in case of amorphization, temperature dependencies of the cumulants do not demonstrate any kinks just like for structural parameters shown in Fig.~\ref{fig:RDF_ind}.

Fig.~\ref{fig:cum} shows concentration dependencies of the cumulants for the initial ($T=1800$ K) and the final ($T=300$ K) states of the Ni-Zr system. The $C(\alpha)$ curves for final ($T=300$ K) states demonstrate discontinuities at the $\alpha=0.1$ and $\alpha=0.66$ which are the boundary concentrations separating regions corresponding to crystalline-like and glassy states (compare with Fig.~\ref{fig:color_map}).

Important result following from the Fig.~\ref{fig:q4-q6_T300},  Fig~\ref{fig:pdf_q6} and Fig.~\ref{fig:cum} is the lack of icosahedral ordering and so the local structure of Ni-Zr alloys is essentially different from that for Cu-Zr ones. To better illustrate this result, we show in Fig.~\ref{fig:ico_vs_x} the comparison of concentration dependencies of icosahedral clusters abundances $n_{\rm ico}$ for Cu-Zr (see Ref.~\cite{Klumov2016JETPLett} for details) and Ni-Zr systems.  A cluster is treated as icosahedral-like one if it has order parameters $q_6 > 0.6$ and $w_6 < -0.16$. It is clearly seen that $n_{\rm ico}$ for Cu-Zr is a few orders of magnitude higher than that for Ni-Zr one. Note that the only composition range corresponding to glassy state for both systems is reasonable to compare ($0.1<x<0.66$ in our case). Note that the conclusion about the lack of icosahedral ordering in Ni-Zr alloys is in close agreement with the results obtained by neutron scattering \cite{Holland-Moritz2009PRB,Fukunaga2006Intermet}, X-ray diffraction \cite{Hao2009PRB} and ab-initio molecular dynamics \cite{Huang2011PRB}.

\section{Discussion}

\begin{figure*}[t]
\centering
\includegraphics[width=0.99\textwidth]{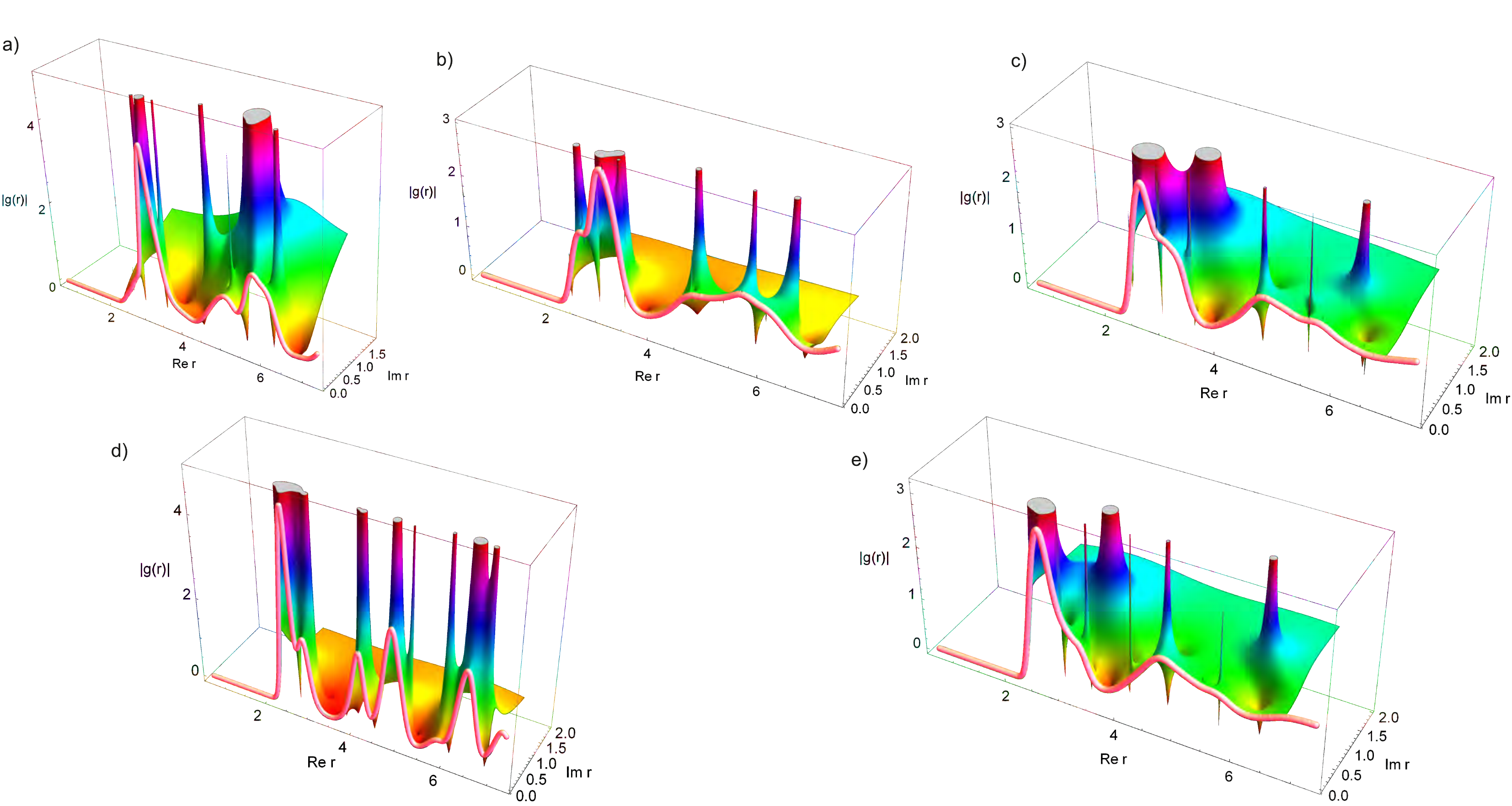}\\
\caption{(Color online)  Numerical analytical continuation of RDF into complex $r$-plane for ${\rm Ni_{\alpha}Zr_{1-\alpha}}$ alloy. Figure (a) and (d) corresponds to crystal state of ${\rm Ni_{10}Zr_{90}}$ and ${\rm Ni_{80}Zr_{20}}$ while (b), (c) and (d) to glass one of ${\rm Ni_{20}Zr_{80}}$, ${\rm Ni_{40}Zr_{60}}$ and ${\rm Ni_{50}Zr_{50}}$. The temperature $T=300$K. The red ``tube'' curve is the original RDF. The ultranarrow peaks are the so called Pade ``defects'' related to some inaccuracy noise in the data-tables of the original RDFs.}
\label{fig:RDF_PADE}
\end{figure*}

\subsection{PADE analysis of two-point correlation functions}

Doing molecular dynamic simulations and growing the crystalline sample from the melt one usually gets a polycrystal consisting of several crystalline grains. The larger the system size is the more grains form in the polycrystalline sample.  So an important question appears how to distinguish the glassy state from the polycrystalline one. From the first glance, it appears to be not very convincing using the radial distribution function since its peaks in the glass and (poly)crystaline states may look quite similar (compare, for example, panels a and b in Fig.~\ref{fig:RDF_tot_vs_T}). So we proposed above an alternative method based on the specific study of the $\phi-\theta$ angular correlations of the nearest neighbors in the spherical coordinate system. For a relatively small system, as we have here, this method proves to be rather effective (see Fig.~\ref{fig:phi-theta}). But, for a large polycrystalline system this method should loose its effectiveness. This is so because symmetry directions of different crystallites are in general quite differently oriented. So the  $z$ axis, pinned to the simulation box, of the spherical coordinate system is also differently oriented with respect to the crystal symmetry directions in different crystallites. Then most of the crystal-like characteristics of the angular distribution should become washed out after averaging over all crystallites of the polycrystal.

Next, it is important that the radial distribution function can be extracted from experimental diffraction data while the $\phi-\theta$ angular distribution can basically not. So this is a fundamental problem how to distinguish between glassy state and polycrystal one by means of an additional special treatment of the radial distribution function.  For that purpose, we propose using the method of numerical analytical continuation into complex $r$-plane (we treat $r$ as the complex number) developed earlier in~\cite{Chtchelkatchev2015JETPLett,Chtchelkatchev2016JETPLett,Khusnutdinoff2016JETP} to process the RDF-functions and uncover its hidden structure. Technically this procedure is based on Pade-approximants built on top of RDF$(r)$ table provided by our computer simulations.

Processed RDFs are shown in Fig.~\ref{fig:RDF_PADE}. As follows, RDF peaks and subpeaks transfer into pole-singularities in the complex plane. That means that RDF's fitting functions should be based on Lorentzians, $\sim 1/[(r-r_0)^2+\gamma^2]$, where each  pole-singularity generates the Lorenzian. Namely, the pole at complex point $z_0$ is related to the term $\propto 1/(r-z_0)$. At real $r$ it reduces to the Lorentzian-like function $\propto 1/[(r-\mathrm{Re}\, z_0)^2+(\mathrm{Im}\, z_0)^2]$. So if only one pole corresponds to the peak, calculation of $\mathrm{Im}\, z_0$ gives the width of the peak. Doing the Fourier transform of RDF one produces the static structure factor $S(q)$. Then the pole singularity in RDF will correspond to terms in $S(q)$ like $\sim\cos(r/\mathrm{Re} z_0) \exp(-r/\mathrm{Im}z_0)$ according to contour integration rules and the residue theorem.

This result with Lorentzians is slightly counterintuitive because RDF is some kind of probability density built on top of large amount of particles interacting non-trivially. So intuitively, keeping in mind the statistics foundations, the Central Limit Theorem, one should naively expect Gaussian peaks. But instead nature produces Lorentzians. For many other systems we have seen the same behaviour of RDF before~\cite{Chtchelkatchev2016JETPLett,Khusnutdinoff2016JETP}. This point with Lorentzians certainly requires further investigations for understanding that will be done later elsewhere.

Important point that complex analytical continuation helps to distinguish glass RDFs from (poly)crystal ones.  Crystal state is characterised by the long range order. For perfect crystal, RDF shows pronounced peaks for first, second, etc. nearest neighbour shells. For the imperfect crystal, the peaks in RDF are broadened and their amplitude is reduced, see, e.g.,  Fig.~\ref{fig:RDF_PADE}(d). In glassy state there is only short-range local order. Then usually only the first and second peaks of RDF (corresponding to first and second shells) are pronounced. Looking, e.g., at original RDFs (red ``tube'' curves) in Figs.~\ref{fig:RDF_PADE}, how to distinguish crystal from glass? The answer will give the second shell peak of RDF. Fig.~\ref{fig:RDF_PADE}(d) clearly shows the crystal, but why Fig.~\ref{fig:RDF_PADE}(a) is crystal? For crystal, Fig.~\ref{fig:RDF_PADE}(a), analytical continuation shows the pronounced pole-singularity with the same amplitude as we see for the first shell of RDF. This is the signature of the long range order. The pole corresponds to characteristic scale, the amplitude of the peak --- to the ``weight'' of this scale among the others. The weight of the pole for second peak is the same like for the first one if we deal with crystal. For glass the situation is different, see, e.g.,  Fig.~\ref{fig:RDF_PADE}(b). The second shell of RDF produces poles with much smaller weight than for the first shell. For glass, we repeat, this is natural since there is no long range order.

\subsection{Glass-forming ability of Ni-Zr alloys}

The results presented above show that simulated Ni-Zr alloys demonstrate either crystal or glassy solid phases being quenched from a liquid state at different cooling rates. So it is interesting to compare GFA of the Ni-Zr system with that for other binary mixtures. The first obvious system for comparison is the chemically similar Cu-Zr one that is a representative model glassformer. Experimentally, the Ni-Zr system has worse GFA than Cu-Zr one \cite{Dong1981JNCS,Huang2013JM}. According to available data, the same is true for simulated Cu-Zr and Ni-Zr alloys.  For widely accepted and intensively studied Cu-Zr EAM potential \cite{Mendelev2009PhilMag,Zhang2015PRB}, it was shown that binary $\textrm{Cu}_{\alpha}\textrm{Zr}_{1-\alpha}$ alloys quenched at cooling rate of $10^{11}$ K/s do not crystallizes in the concentration range of $\alpha\in(0.2-0.95)$ \cite{Klumov2016JETPLett}. Long time molecular dynamic simulations of binary $\textrm{Cu}{64.5}\textrm{Zr}_{35.5}$ alloy shows that the system partially crystallizes at cooling rate of the order of $10^9$ K/s \cite{Ryltsev2016JCP}.

Another representative binary system is a binary Lennard-Jones mixture (BLJ). There are two widely accepted BLJ glassformers -- Kob-Andersen \cite{Kob1994PRL} and Wahnstr\"om ones \cite{Wahnstrom1991PRA} -- which was specially designed to favor GFA. For example, interaction parameters of Kob-Andersen model was chosen to mimic strong chemical interaction between components (that means non-additive
mixture
$\sigma_{\scriptscriptstyle{\rm AB}}\neq(\sigma_{\scriptscriptstyle{\rm AA}} +\sigma_{\scriptscriptstyle{\rm BB}})/2$ with $\epsilon_{\scriptscriptstyle{\rm AA}}\neq\epsilon_{\scriptscriptstyle{\rm BB}}\neq\epsilon_{\scriptscriptstyle{\rm BB}}$). These models are usually considered at certain compositions (${\rm A_{80}B_{20}}$ and ${\rm A_{50}B_{50}}$, respectively) and so a direct comparison with $\textrm{Cu}_{\alpha}\textrm{Zr}_{1-\alpha}$ alloys is hardly realizable. Note that both Kob-Andersen and Wahnstr\"om models have been recently revealed as not so good glassformers because they partially crystallized in lengthy molecular dynamics simulations \cite{Toxvaerd2009JCP}. Moreover, Kob-Andersen mixture is in fact very poor glassformer for large ($N>10000$) system sizes \cite{Ingebrigtsen2018crystallisation}.

For our purposes, the more interesting results were obtained in \cite{Shimono2001ScriptaMater,Shimono2012RevueMet} where comprehensive study of GFA of  additive BLJ without chemical interaction (i.e. $\sigma_{\scriptscriptstyle{\rm AB}}=(\sigma_{\scriptscriptstyle{\rm AA}} +\sigma_{\scriptscriptstyle{\rm BB}})/2$ and $\epsilon_{\scriptscriptstyle{\rm AA}}=\epsilon_{\scriptscriptstyle{\rm BB}}=\epsilon_{\scriptscriptstyle{\rm BB}}$)
was performed. Particularly, the concentration regions corresponding to both crystal and glassy states were obtained at different cooling rates and atomic size ratios. Let us consider formal accordance between parameters of BLJ studied in \cite{Shimono2001ScriptaMater,Shimono2012RevueMet} and Ni-Zr alloys under consideration. The analysis of partial RDFs (Fig.~\ref{fig:RDF_part_T300}) allows estimating the atomic size ratio for Ni-Zr system as
$r_{\rm Ni}/r_{\rm Zr}\approx 0.8$. The cooling rate range of $10^{10}-10^{13}$ K/s corresponds to the range of $5\cdot 10^{-6}-2\cdot 10^{-3}$ in reduced Jennard-Jones units used
in\cite{Shimono2001ScriptaMater,Shimono2012RevueMet}. Doing such parameter matching, we conclude that concentration ranges of glass-formation for Ni-Zr system are approximately the same as those for BLJ one. That means GFA of EAM Ni-Zr model considered is the same order as for additive binary mixture of atoms interacting by isotropic pair potential without any chemical interaction. This conclusion is in close agrement with the lack of icosahedral ordering observed for the system under investigation (see Fig.~\ref{fig:ico_vs_x}). So we suggest that peculiarities of interatomic interactions in the considered EAM model of Ni-Zr alloys do not favor the formation of local icosahedral clusters and that is the reason of poor GFA of the system.

\section{Conclusions}

Doing molecular dynamic simulations and quenching the Ni-Zr system at different cooling rates, we observe that final room temperature state is a glass in certain (cooling rate dependent) concentration range $\alpha_{\rm min}<\alpha<\alpha_{\rm max}$. Comparing such "glassy window" with those for both Cu-Zr  system and binary Lennard-Jones mixture we conclude the GFA of Ni-Zr  system is nearly the same as for the latter.  We suggest that such relatively weak GFA of the system under consideration is due to the fact that its local structure does not contain noticeable amount of icosahedral clusters. This result supports widely accepted idea that poly-tetrahedral (icosahedral) structural motifs favor glass-forming ability of particle systems.

The Cu-Zr and Ni-Zr systems are similar in physicochemical properties of component elements. But, replacing Cu with Ni, we change the ratio between interparticle scales, which seriously affects the frustration in the systems. The ratio between scales in the Cu-Zr system stabilizes the icosahedra, while in the Ni-Zr system -- it does not.

\section{Acknowledgments}

MD Simulations were performed within the framework of state assignment of IMET UB RAS (the project №0396-2015-0076). Structural analysis was supported by Russian Science Foundation (grant grant RNF 18-12-00438). Analysis of Cu-Zr system was supported by Russian Science Foundation (grant   №14-13-00676). Institute of Metallurgy thanks UB RAS for access to "Uran" cluster. Part of calculations was performed using the resources of the Federal Collective Usage Center Complex for Simulation and Data Processing for Mega-science Facilities at NRC ``Kurchatov Institute'', http://ckp.nrcki.ru/, and  the cluster of Joint Supercomputing Center, Russian Academy of Sciences.

\bibliography{bib_glasses}
\end{document}